\documentclass{jfm}
\usepackage{graphicx}
\usepackage{epstopdf, epsfig}

\newcommand\Ray{\mbox{\textit{Ra}}}  % Rayleigh number
\newcommand\Nus{\mbox{\textit{Nu}}}  % Nusselt number
\newcommand\Ek{\mbox{\textit{E}}}  % Ekman number
\newcommand\ri{r_{\rm i}} % inner radius
\newcommand\ro{r_{\rm o}} % outer radius
\newcommand\Ros{\mbox{\textit{Ro}}}  % Rossby number

\usepackage{color}

\shorttitle{Heat Flow and Boundary Heterogeneity in Rotating Convection}
\shortauthor{J. E. Mound and C. J. Davies}

\title{Heat Transfer in Rapidly Rotating Convection with Heterogeneous Thermal Boundary Conditions}

\author{Jon E. Mound\aff{1}
  \corresp{\email{j.e.mound@leeds.ac.uk}},
 \and Christopher J. Davies\aff{1,2}}

\affiliation{\aff{1}School of Earth and Environment, University of Leeds, Leeds LS2 9JT, UK
\aff{2} Institute of Geophysics and Planetary Physics, Scripps Institution of Oceanography,
University of California at San Diego, 9500 Gilman Drive, La Jolla, CA 92093-0225, USA
}

\begin{document}

\maketitle

\begin{abstract}
Convection in the metallic cores of terrestrial planets is likely to be subjected to lateral variations in heat 
flux through the outer boundary imposed by creeping flow in the overlying silicate mantles. Boundary anomalies can 
significantly influence global diagnostics of core convection when the Rayleigh number, $\Ray$, is weakly supercritical; 
however, little is known about the strongly supercritical regime appropriate for planets. We perform numerical simulations
of rapidly rotating convection in a spherical shell geometry and
impose two patterns of boundary heat flow heterogeneity: a hemispherical $Y_1^1$ 
spherical harmonic pattern; and one derived from seismic tomography of Earth's lower mantle. We 
consider Ekman numbers $10^{-4} \le \Ek \le 10^{-6}$, flux-based Rayleigh numbers up to $\sim800$ times critical, 
and Prandtl number unity. The amplitude of the lateral variation in heat flux is characterised by 
$q_{\rm L}^*=0, 2.3, 5.0$, the peak-to-peak amplitude of the outer boundary heat flux divided by its mean. 
We find that the Nusselt number, $\Nus$, can be increased by up to $\sim25$\% relative to the equivalent homogeneous case 
due to boundary-induced correlations between the radial velocity and temperature anomalies near the top of the shell.
The $\Nus$ enhancement tends to become greater as the amplitude and
length scale of the boundary heterogeneity are increased and as the system becomes more supercritical.  
This $\Ray$ dependence can steepen the $\Nus \propto \Ray^\gamma$ scaling
in the rotationally dominated regime, with $\gamma$ for our most extreme case approximately $20\%$ 
greater than the equivalent homogeneous scaling. 
Therefore, it may be important to consider boundary heterogeneity when extrapolating numerical 
results to planetary conditions.
\end{abstract}

\begin{keywords}
Authors should not enter keywords on the manuscript, as these must be chosen by the author during the online submission process and will then be added during the typesetting process (see http://journals.cambridge.org/data/\linebreak[3]relatedlink/jfm-\linebreak[3]keywords.pdf for the full list)
\end{keywords}

\section{Introduction}

Convection arises in many natural systems, from oceans and atmospheres to terrestrial mantles 
and liquid metal planetary cores. These systems often operate in dynamical regimes that cannot be 
attained in current laboratory experiments or numerical models. Significant effort has therefore 
focused on elucidating scaling relationships between the independent variables and diagnostics of the 
system behaviour. A particularly important diagnostic is the Nusselt number, $\Nus$, a global 
measure of the efficiency of heat transport in the convecting system. 
This study is motivated by convection in low viscosity, liquid metal, planetary cores where rotation and the spherical 
geometry significantly affect the dynamics and convection in 
the overlying silicate mantles sets the thermal boundary conditions at the top of core.
Lateral variations in the temperature field at the bottom of silicate mantles can be very large
relative to those expected in the metallic cores, resulting in correspondingly large lateral
variations in the thermal boundary conditions imposed on the underlying core.
 We focus here on the impact of such heterogeneous boundary 
conditions on the heat transfer behaviour of rapidly rotating convection in spherical geometry.

\subsection{Heat transport in convecting systems}

The canonical system used to understand convective heat transfer is a conducting fluid sandwiched 
between two parallel plates oriented normal to the gravity vector, with an imposed temperature difference 
$\Delta T$ across the layer. The system is characterised by the Rayleigh number, $\Ray$, 
measuring the strength of the convective driving and the Prandtl number, $\Pran = \nu/\kappa$, where 
$\nu$ is the kinematic viscosity and $\kappa$ the thermal diffusivity. For a given value of $\Pran$, as $\Ray$ is increased more 
heat is transported by advection and so $\Nus = \epsilon \Ray^{\gamma}$, with $\epsilon, \gamma > 0$. 
\citet{Malkus:1954wq} assumed that $\Delta T$ is accommodated predominantly by conduction in two identical thermal 
boundary layers at the top and bottom of the system; by further assuming that the local Rayleigh number of the boundary layer 
equals the critical value for stability he found that $\Nus \propto \Ray^{1/3}$.  
Numerical and physical experiments have produced a variety of scalings, which also depend on $\Pran$.
 \citet{Grossmann:2000dc} \citep[see also][]{Grossmann:2001hk, Grossmann:2002ir, Stevens:2013gg} 
 argued that these differences depend on whether the 
 boundary layers or the bulk of the fluid dominate the kinematic and thermal dissipations, which divides
  $\Ray$-$\Pran$ space into four regimes, each of which is divided into two 
 sub-regimes based on the relative thickness of the thermal and kinetic boundary layers. 

When the convecting system is rotating an additional nondimensional parameter, 
the Ekman number, $E$, quantifies the relative strength of the viscous and Coriolis forces. 
The Coriolis force has a stabilising effect on the convection such that the critical Rayleigh number for the
onset of convection, $\Ray_{\rm C}$, depends on $E$. 
When convection is geostrophic, that is the first order force balance is between the Coriolis force and
pressure gradients, a heat transport scaling of  $\Nus \propto \Ray^3 \Ek^4$ has been proposed \citep{King:2012gi} 
following the same reasoning as \citet{Malkus:1954wq}. 
However, in rapidly rotating systems a substantial interior temperature gradient can be maintained even to high $\Ray$
\citep{Sprague:2006js, Julien:2012hs, King:2013fx, Gastine:2016gq, Julien:2016js}, which enhances
diffusive heat transport and alters $\Nus$ relative to an equivalent non-rotating case. 
Experiments and numerical simulations in Cartesian geometry suggest that $\gamma$ increases as $E$ decreases into the
rotationally-dominated regime \citep{Stellmach:2014kp, Cheng:2015ka}.  
At $E=10^{-7}$ a value of $\gamma=3.6$ has been found, with no indication
that the scaling had yet reached an asymptotic limit with reducing $\Ek$ \citep{Cheng:2015ka}. Reduced equations valid in the 
asymptotic limit of small $E$ \citep{Sprague:2006js} also find $\gamma>3$ at low $\Ek$ when the effect of Ekman pumping from
 no-slip boundary conditions is included \citep{Stellmach:2014kp, Aurnou:2015jh, Julien:2016js, Plumley:2016eo}.
Conversely, with free-slip boundary conditions the scaling at
low-$\Ek$ appears to saturate with $\gamma = 3/2$, which is expected
for a turbulent quasi-geostrophic regime where the heat transport is independent of
both the thermal and viscous diffusivities \citep{Gillet:2006be, Julien:2012dc, Stellmach:2014kp}.

In the spherical geometry considered in this paper, the boundary layer analysis is further complicated by asymmetric
boundary layers, with the asymmetry in boundary layer thicknesses and temperature drops
depending on both the radius ratio between the inner and outer boundaries and on the radial dependence of gravitational acceleration
within the shell \citep{Gastine:2015cj}.  In the absence of rotation the $\Nus$-$\Ray$ scaling in the shell is similar to that of the plane layer 
\citep{Gastine:2015cj}. At sufficiently high $\Ray$ buoyancy forces will dominate
Coriolis forces resulting in effectively non-rotating dynamics and $\Nus$-$\Ray$ scaling.
In the presence of rotation \citet{Gastine:2016gq} obtained the diffusivity-free scaling $\Nus \propto \Ray^{3/2}\Ek^2$ in a small region
of parameter space with
$\Ek \lesssim 10^{-6}$ and $\Ray\Ek^{4/3} \simeq 10$, conditions that placed the simulations in a dynamical regime
that was both strongly non-linear and dominated by rotation.

\subsection{Heterogeneous boundary conditions for the core}

Liquid metal planetary cores have higher thermal conductivity, and much lower viscosity, than their overlying solid silicate
mantles. The resultant asymmetry in the thermal evolution
of these systems implies that when considering thermal core-mantle 
interaction in simulations of core convection the use of fixed-flux boundary conditions would apply, 
whereas fixed-temperature boundary conditions would apply for the mantle \citep{Olson:2003wy, Olson:2016ea}. 
Choosing fixed-temperature or fixed-flux boundary conditions results in different formulations of the Rayleigh and Nusselt numbers;
these differences are outlined in \S\ref{sec:theory}.
However, after appropriate translation between the
fixed-flux and fixed-temperature formulations the Nusselt-Rayleigh scaling is the 
same in both cases \citep{Otero:2002by, Ahlers:2009hx, Johnston:2009fx, Calkins:2015gi, Goluskin:2015iw}. 

\begin{figure}
  \centerline{\includegraphics{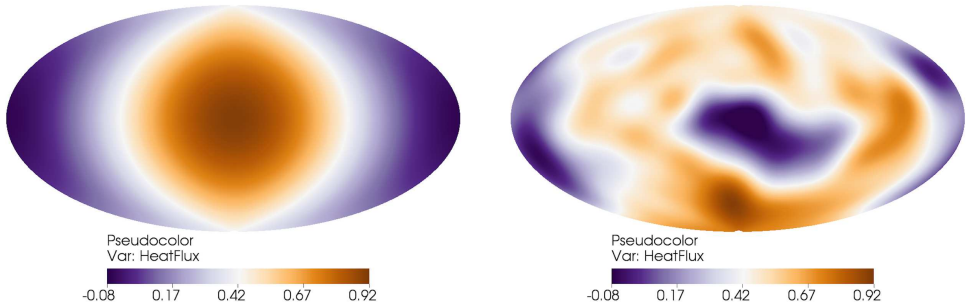}}% Images in 100% size
  \caption{Patterns of core-mantle boundary heat flux with hemispheric (left) and tomographic (right) perturbations added to the mean. 
  The projection is centred on the negative $x$-axis (the Pacific). In both cases
  $q_{\rm L}^\star = 2.3$ and 
  the choice of normalisation results in a mean heat flux of approximately 0.42; 
  note that only the deepest purples are associated with a negative (i.e.\ radially inward) heat flux.}
\label{fig:Patterns}
\end{figure}

Mantle convection is much slower and supports much larger lateral variations in fluid properties 
(e.g. density, temperature, composition) than core convection \citep{Olson:2003wy, Olson:2016ea}. It is 
therefore widely believed that convection in planetary cores must respond to a laterally-varying pattern 
of heat flow imposed at the core-mantle boundary by mantle convection \citep{Amit:2015fh}. 
For the present-day Earth, a pattern of heat flux (right panel of figure~\ref{fig:Patterns}) is suggested by seismic tomography \citep[e.g.][]{Masters:1996ey}, 
which reveals two large low shear velocity provinces (LLSVPs) at the base of the mantle interpreted as hot, dense
thermochemical piles in roughly antipodal locations beneath the Pacific Ocean and Africa \citep{Garnero:2016dz}. Seismically 
fast material between these LLSVPs is interpreted as the cold remnant of subducted lithospheric slabs. The largest component of 
this tomographic pattern is spherical harmonic of degree and order two ($Y_2^2$), although other components also contribute.
Inclusion of this pattern of heat flux in numerical models of the geodynamo can produce features in
the spatial structure and temporal variation of the magnetic field similar to those observed for the Earth 
\citep{Bloxham:2000uu, Olson:2002tk, Gubbins:2007cn, Willis:2007cu, Davies:2008jx, Amit:2010ks, Amit:2015bc, Olson:2016ea}.

The present-day Earth is but one example of the pattern, and amplitude, of
core-mantle boundary heat flux heterogeneity that can arise in planetary mantle convection. During the assembly of super-continents 
a primarily hemispheric ($Y_1^1$) pattern of mantle convection and hence core-mantle boundary heat flux (left panel of figure~\ref{fig:Patterns}) 
may have existed \citep{Zhong:2007ev, Zhang:2011bz, Olson:2013fd, Olson:2016ea}. 
A hemispheric heat flux pattern has also been suggested for Mars to explain the Tharsis bulge \citep{Zhong:2009bg, Sramek:2010dh},
for the Moon to generate a non-axial magnetic field \citep{Takahashi:2009jy, Oliveira:2017bm},
and for hot tidally-locked terrestrial exoplanets \citep{Gelman:2011ff}. 

The amplitude of the lateral variations in the heat flux conducted through the outer boundary can be expressed as
\begin{equation}
q_{\rm L}^\star = \frac{q_{\rm max} - q_{\rm min}}{q_{\rm ave} - q_{\rm ad}}, 
\label{eq:qstardef}
\end{equation}
where $q_{\rm max}$,  $q_{\rm min}$, and $q_{\rm ave}$  are the maximum, minimum, and horizontally averaged heat flux through
the boundary, respectively, and $q_{\rm ad}$ is the heat flux conducted down the adiabatic gradient of the core
at the boundary. $q_{\rm L}^\star$ is challenging to precisely determine for Earth because it depends on 
the convective fluctuations as well as material properties of the mantle; it also will have varied through time. 
Estimates of the present-day total heat flow across the core-mantle boundary generally fall in the range 
$Q_{\rm CMB} =  5$--20~TW \citep{Lay:2008wb, Nimmo:2015fy, Kavner:2016wb}. 
Mantle convection simulations predict $(q_{\rm max} - q_{\rm min}) / q_{\rm ave} = {\it O}(1)$ for Earth 
\citep{Nakagawa:2008kk, Olson:2015kx}. Recent upward revisions of the thermal conductivity of liquid 
iron mixtures increase $q_{\rm ad}$ so that it is comparable to estimates of $q_{\rm ave}$ \citep{Lay:2008wb, Davies:2015bo}, 
in which case $q_{\rm L}^\star \ge {\it O}(1)$. 
The lateral variations that we include in our model all have zero mean, therefore values of $q_{\rm L}^\star > 2$ 
imply that for some portion of the outer boundary the heat flux is radially inward, although the integrated flux would remain outward. 
The average heat flux through the core-mantle boundary enters into the flux Rayleigh number for the core, 
for each flux Rayleigh number we will have a set of homogeneous and heterogeneous cases. 
 Examples of the heat flux through the core-mantle boundary with $q_{\rm L}^\star = 2.3$
are shown in figure~\ref{fig:Patterns} for both hemispheric and tomographic perturbations to the mean.

Numerical \citep{Zhang:1993ts, Zhang:1996ck, Gibbons:2007ef, Davies:2009il, Dietrich:2016ba} 
and physical \citep{Sumita:1999bg, Sumita:2002ce} experiments have investigated rotating convection
with laterally varying thermal boundary conditions for a variety of imposed patterns and 
$q_{\rm L}^\star \ll 1$ up to $q_{\rm L}^\star = {\it O}(1)$. The previous numerical work, conducted predominantly 
at $\Ray$ a few times critical, has shown that imposed boundary heterogeneity can have a substantial 
influence on the spatial patterns and time-dependence of convection in the fluid shell, particularly 
when the wavelength of the imposed pattern is large. However, to our knowledge, the only previous 
study that compared heat transfer behaviour in rotating spherical shells with and without heterogeneous 
outer boundary forcing was that of \citet{Dietrich:2016ba} who carried out simulations 
of internally heated rotating spherical shell convection with a $Y_1^1$ outer boundary heat flux pattern 
at $\Pran=1$ and $E=2.5, 5\times 10^{-5}$ (using the definition of $E$ in \S\ref{sec:Governing}) and flux-based Rayleigh numbers
 up to 400 times critical for $E=5\times 10^{-5}$. Estimating $\Delta T$ based on the maximum and minimum 
values of $T$ in the domain \citet{Dietrich:2016ba} found that $\Nus$ is reduced by up to 50\% from the homogeneous case
as the amplitude of heterogeneity is increased up to $q_{\rm L}^{\star}=2$ (using our definition~\ref{eq:qstardef}).

Here we present results from 106 numerical simulations of bottom-heated, rotating convection in a spherical shell 
with values of $E$ as low as $10^{-6}$. We include two patterns of boundary heterogeneity and extend 
to flux-based Rayleigh numbers several hundred times critical. We focus in particular on heat 
transport within these models and the impact of the boundary heterogeneity on $\Nus$ as a measure of 
convective efficiency. In \S\ref{sec:theory} we outline our theoretical basis for exploring heat transport 
in rotating convection; in \S\ref{sec:results} we present summary results of all of our numerical
simulations and more extended discussion of certain illustrative cases.
 
\section{Theory}\label{sec:theory}

\subsection{Governing equations and non-dimensionalisation}\label{sec:Governing}
We employ a numerical model of convection of a homogeneous Boussinesq fluid confined within a rotating spherical shell
\citep{Willis:2007cu}. The fluid is characterised by its constant thermal diffusivity, $\kappa$,  kinematic viscosity, $\nu$,
coefficient of thermal expansion, $\alpha$, and reference density, $\rho_0$. 
The thermal diffusivity can be expressed as $\kappa = k / \rho_0 C_P$, where $k$ is the thermal conductivity and $C_P$ the heat capacity
of the fluid.
The shell is defined in spherical coordinates, $(r,\theta,\phi)$, 
by the inner and outer boundaries, $\ri$ and $\ro$, respectively, and rotates with a constant angular velocity
$\boldsymbol{\Omega} = \Omega\boldsymbol{\hat{z}}$.
The governing equations  for conservation of momentum, energy, and mass can be written

\begin{equation}
 \left( \frac{\partial\boldsymbol{u} }{ \partial t} + (\boldsymbol{u\cdot\nabla})\boldsymbol{u} \right)
+ 2\boldsymbol{\Omega\times u} = -\frac{1}{\rho_0}\nabla \widetilde{P} + \frac{\rho}{\rho_0}\boldsymbol{g} + \nu\nabla^2\boldsymbol{u},
\label{eq:NavierStokesDim}
\end{equation}

\begin{equation}
\frac{\partial T}{\partial t} + \boldsymbol{\left( u\cdot\nabla\right) }T = \kappa \nabla^2T,
\label{eq:EnergyTDim}
\end{equation}

\begin{equation}
\boldsymbol{\nabla\cdot u} = 0.
\label{eq:MassDim}
\end{equation}
The modified pressure, $\widetilde{P}$, includes the centrifugal potential. 
Gravity varies linearly with radius such that
$\boldsymbol{g} = -(g_{\rm o}/\ro)\boldsymbol{r}$, where $g_{\rm o}$ is the gravitational acceleration at $r=\ro$.

The fluid is incompressible (\ref{eq:MassDim}) and so the velocity, $\boldsymbol{u}$, can be decomposed 
into toroidal, $\mathcal{T}$, and poloidal, $\mathcal{P}$, components such that
\begin{equation}
\boldsymbol{u} = \nabla\times\left(\mathcal{T}\mathbf{\hat{r}}\right) + \nabla\times\nabla\times\left(\mathcal{P}\mathbf{\hat{r}}\right).
\label{eq:VelTorPolDecomp}
\end{equation}The toroidal and poloidal scalar fields can then be expressed in terms of spherical harmonics, $Y_\ell^m$, 
with radially varying harmonic coefficients $\tau_\ell^m(r)$ and $p_\ell^m(r)$ respectively. In this work we make use of the
Schmidt semi-normalised spherical harmonics common in geomagnetic studies \citep[e.g.][]{Kono:2015eu}.
The boundary conditions on the velocity are non-penetrative and no-slip such that $\boldsymbol{u} =0$ on $r=\ri,\ro$.

The temperature field, $T$, can be written as 
\begin{equation}
T=T_{\rm C} + T^\prime,
\label{eq:Tdecomp}
\end{equation}
where $T_{\rm C}$ is the steady-state temperature in the absence of flow and
$T^\prime$ is the fluctuation about this state. We denote the coefficients of the spherical harmonic expansion of $T$ by $\vartheta_\ell^m(r)$.
Fixed-flux thermal boundary conditions are imposed such that $\boldsymbol{\nabla} T_{\rm C} = -(\beta/r^2)\boldsymbol{\hat{r}}$ at
the inner and outer boundaries. Thus the total heat flow, $Q$, is equal through the inner and outer surfaces and,
for example, on the outer boundary
\begin{equation}
Q = 4 \pi \ro^2 \, q_{\rm ave} = 4 \pi \ro^2 \left(-k \boldsymbol{\nabla} T_{\rm C} \right) = 4\pi k \beta.
\label{eq:QtoBeta}
\end{equation}

We introduce the following notation for radial, spherical surface, and time averages, respectively
\begin{eqnarray}
\left\{ f(r) \right\} &=& \frac{1}{h} \int_{\ri}^{\ro} f(r)\:\mathrm{d}r, \\
\langle f(r,\theta,\phi) \rangle &=& \frac{1}{4\pi r^2} \int_0^\pi \int_0^{2\pi} f(r,\theta,\phi)\; r^2\sin\theta \:\mathrm{d}\phi \:\mathrm{d}\theta,  \\ 
 \overline{f(t)} &=& \frac{1}{\tau} \int_{t_0}^{t_0 + \tau} f(t) \:\mathrm{d}t, 
\end{eqnarray}
where $h = \ro - \ri$ is the shell thickness and $\tau$ is the duration of the time averaging.

The control parameters that characterise the convecting system are derived from non-dimensionalisation of the governing
equations (\ref{eq:NavierStokesDim})--(\ref{eq:MassDim}).
We scale length by the shell thickness, $h$, time by the thermal diffusion time, $\tau_{\rm d} = h^2/\kappa$,
and temperature by $\beta/h$.
With this choice of scaling the Ekman number, Prandtl number, and modified Rayleigh number can be defined as
\begin{equation}
\Ek = \frac{\nu}{2\Omega h^2}, \quad
\Pran = \frac{\nu}{\kappa}, \quad
\widetilde{\Ray} = \frac{\alpha g_{\rm o} \beta}{2\Omega\kappa},
\label{eq:NonDimNumbers}
\end{equation}
and the resultant non-dimensional governing equations are
\begin{equation}
\frac{\Ek}{\Pran} \left( \frac{\partial\boldsymbol{u^\star} }{ \partial t^\star} + (\boldsymbol{u^\star\cdot\nabla^\star})\boldsymbol{u^\star} \right)
+ \boldsymbol{\hat{z}\times u^\star} = -\nabla^\star \widetilde{P^\star} + \widetilde{\Ray} T^{\prime\star}\boldsymbol{r^\star} + \Ek\nabla^{\star 2}\boldsymbol{u^\star},
\label{eq:NavierStokes}
\end{equation}

\begin{equation}
\frac{\partial T^\star}{\partial t^\star} + \boldsymbol{\left(u^\star\cdot\nabla^\star\right)}T^\star = \nabla^{\star 2}T^\star,
\label{eq:EnergyT}
\end{equation}

\begin{equation}
\boldsymbol{\nabla^\star\cdot u^\star} = 0.
\label{eq:Mass}
\end{equation}
In all of our models $\Pran = 1$ and the radius
ratio of the shell is set to $\ri/\ro = 0.35$, the value for Earth's core. 
The modified Rayleigh number is related to a flux Rayleigh number by
\begin{equation}
\Ray_{\rm F}  = \frac{\alpha g_{\rm o} \beta h^2}{\nu\kappa} = \frac{ \widetilde{\Ray} }{ \Ek }.
\end{equation}
To convert from flux-based to temperature-based Rayleigh we need to relate $\beta$ to
the temperature drop across the convecting system, $\Delta T$.
The solution to the spherical shell conduction problem 
(see equations~\ref{eq:SphericalDiffusionEq}--\ref{eq:DeltaTCondShell} below)
gives $\beta = \Delta T_{\rm C}\ri\ro/h$, which 
in combination with our expression for the Nusselt number (equation~\ref{eq:NuDef} below) leads to
\begin{equation}
\Ray_{\rm T} = \frac{\alpha g_{\rm o}  \Delta T h^3}{\nu\kappa} = \frac{\Ray_{\rm F}}{\Nus}\frac{h^2}{\ri\ro},
\end{equation}
where $\Delta T$ is taken to be $ \overline{\Delta\langle T\rangle}$, the measured time-averaged temperature drop 
across the convecting system. 

Below we will generally present results relative to the advection time scale, $\tau_{\rm a} = h/U$,
where $U$ is a characteristic velocity for the flow.  The ratio between thermal diffusion and advection
time scales is the thermal P\'eclet number,  $\Pen = \tau_{\rm d} / \tau_{\rm a} = U h/ \kappa$. 
 The thermal P\'eclet number is the product of the Prandtl and Reynolds numbers; 
since we set $\Pran=1$ in all of our models we have $\Pen = \Rey = U h/ \nu$.
The characteristic velocity is derived from the kinetic energy and after non-dimensionalisation we have
\begin{equation}
\Pen = \Rey = U^\star = \sqrt{\frac{2{\rm {KE}^\star}}{{V{\rm s}}^\star}},
\label{eq:PeReUDef}
\end{equation} 
where ${V{\rm s}}^\star$ is the non-dimensional volume of the domain and ${\rm {KE}^\star}$ is the non-dimensional kinetic energy
of the system defined by
\begin{equation}
{\rm {KE}^\star} = \frac{1}{2} \int\!\!\!\int\!\!\!\int_{{V{\rm s}}^\star} \overline{\boldsymbol{u^\star}^2} dV^\star.
\label{eq:KEdef}
\end{equation}

\subsection{Heat transport}\label{sec:NusseltTheory}

The Nusselt number provides a global measure
of the efficiency of heat transport in a convecting system by comparing the total
heat flux through the system to that which could be transferred by conduction alone. 
Equation~(\ref{eq:EnergyTDim}) can be written as
\begin{equation}
\rho_0 C_P \frac{\partial T}{\partial t} + \boldsymbol{\nabla\cdot q} = 0,
\label{eq:Energyq}
\end{equation}
where the total heat flux
\begin{equation}
\boldsymbol{q} = \rho_0 C_P \boldsymbol{u}T - k \boldsymbol{\nabla}T
\label{eq:qtotaldef}
\end{equation}
 is the sum of the advective and diffusive contributions.

We first revisit the canonical example of a plane layer of thickness of $d$ and  fixed temperature difference 
across the layer of $\Delta T$. In
this case the Nusselt number is
 \begin{equation}
\Nus = \frac{ \langle { \boldsymbol{ \overline{q} \cdot \hat{z}}  \rangle} }{k \Delta T/d}, 
\label{eq:NuSimpleDef}
\end{equation}
where the total vertical heat flux, $ \boldsymbol{q \cdot \hat{z} } $, can be averaged over any horizontal surface
and the conduction solution gives $(\boldsymbol{\nabla}T_C)\boldsymbol{\cdot\hat{z}} = -\Delta T/d$.
It can be particularly useful to consider the top or bottom surface, where $ \boldsymbol{q \cdot \hat{z} } $ is purely conductive,
such that the Nusselt number can be written
\begin{equation}
\Nus  = \frac{ - \langle ( \boldsymbol{\nabla}\overline{T})\boldsymbol{\cdot\hat{n}} \rangle|_{\rm top} }
	{ - (\boldsymbol{\nabla}T_C)\!\boldsymbol{\cdot\hat{n}}|_{\rm top} }
= \frac{  \langle (\boldsymbol{\nabla}\overline{T})\boldsymbol{\cdot\hat{n} \rangle}|_{\rm bottom} }
	{  (\boldsymbol{\nabla}T_C)\!\boldsymbol{\cdot\hat{n}}|_{\rm bottom} },
\label{eq:NuGradTs}
\end{equation}
where $\boldsymbol{\hat{n}}$ is the outward normal with respect to the domain. 

In numerical or physical experiments with fixed-temperature boundary conditions
the Nusselt number can thus be evaluated by determining $\langle  \boldsymbol{\overline{q} \cdot \hat{n} } \rangle$ 
for a sufficiently large averaging time.
Fixed-flux boundary conditions set $\boldsymbol{\nabla} T = \boldsymbol{\nabla}T_C$ on both boundaries at every instant in time, in
which case (\ref{eq:NuGradTs}) suggests $\Nus =1$ regardless of convective vigour.
Although the non-penetration condition requires that all heat is transferred by conduction across the boundaries, 
within the fluid interior the advective heat transport will reduce the temperature gradient. 
In experiments where  $\langle  \boldsymbol{\overline{q} \cdot \hat{n} } \rangle$  is fixed, the temperature drop across the system is
the quantity to be determined in (\ref{eq:NuSimpleDef}). 

In our model, with fixed-flux boundaries and spherical shell geometry, the conduction problem is 
\begin{equation}
\frac{\kappa}{r^2}\frac{\mathrm{d}}{\mathrm{d}r}\left( r^2 \frac{\mathrm{d}T_{\rm C}}{\mathrm{d}r}\right) = 0, 
\label{eq:SphericalDiffusionEq}
\end{equation}
subject to boundary conditions
\begin{equation}
\left. \frac{\mathrm{d}T_{\rm C}}{\mathrm{d}r} \right|_{r=\ro} = -\frac{\beta}{\ro^2}, \quad
 \left. \frac{\mathrm{d}T_{\rm C}}{\mathrm{d}r} \right|_{r=\ri} = -\frac{\beta}{\ri^2}.
\end{equation}
The solution is
\begin{equation}
\frac{\mathrm{d}T_{\rm C}}{\mathrm{d}r} = -\frac{\beta}{r^2}, \quad
T_{\rm C} = \frac{\beta}{r} + B
\label{eq:TCondShell}
\end{equation}
where $B$ is a constant of integration that is not constrained by the flux boundary conditions.
Therefore, the temperature drop across the shell in the conduction only case is
\begin{equation}
\Delta T_{\rm C} = \beta\left(\frac{1}{\ri} - \frac{1}{\ro}\right) = \frac{\beta h}{\ri\ro}. 
\label{eq:DeltaTCondShell}
\end{equation}

The total heat flux of the convecting system is found
by time averaging equation~(\ref{eq:Energyq}) for a duration that is sufficiently long to reach a statistical steady state,
in which case   
$\boldsymbol{\nabla\cdot \overline{q}} = 0$ within the domain, a consequence of the absence of internal heat sources.
It follows that
\begin{equation}
\int\!\!\!\int\!\!\!\int_{V{\rm s}}  \boldsymbol{\nabla\cdot \overline{q}} \:\mathrm{d}V = 0 = \oint \boldsymbol{ \overline{q} \cdot \hat{n}} \:\mathrm{d}S
= 4\pi\ro^2 \langle  \boldsymbol{ \overline{q} \cdot \hat{r}} \rangle |_{r=\ro} - 4\pi\ri^2 \langle  \boldsymbol{ \overline{q} \cdot \hat{r}} \rangle |_{r=\ri}
\end{equation} 
and, since the volume of integration is arbitrary,  
$4\pi r^2 \langle \boldsymbol{\overline{q}\cdot\hat{r}} \rangle$ is independent of $r$.
The imposed boundary conditions require $u_r|_{r=\ri} = u_r|_{r=\ro} = 0$, $\partial\overline{T}/\partial r  |_{r=\ro} = -\beta/\ro^2$,
and $\partial\overline{T}/\partial r  |_{r=\ri} = -\beta/\ri^2$, therefore 
\begin{equation}
4\pi r^2 \left\langle \boldsymbol{\overline{q}\cdot\hat{r}}\right\rangle = 
4\pi\ro^2 \left\langle - k\partial\overline{T}/\partial r \right\rangle |_{r=\ro} = 
4\pi\ri^2 \left\langle - k\partial\overline{T}/\partial r \right\rangle |_{r=\ri}=
4\pi k \beta
\label{eq:qconstraint}
\end{equation}
and
\begin{equation}
\left\langle  \boldsymbol{\overline{q}\cdot\hat{r}} \right\rangle = k\beta/r^2 = -k (\mathrm{d}T_{\rm C}/\mathrm{d}r).
\label{eq:qdotr=dTcdr}
\end{equation}

Although $4\pi r^2 \langle \boldsymbol{\overline{q}\cdot\hat{r}} \rangle$ is independent of $r$, 
the advective and diffusive contributions to the total radial heat flux will vary.
The global balance between the advective and diffusive contributions  
requires integration with respect to $r$, 
leading to the following expression for the Nusselt number based on fluxes
\begin{equation}
\Nus_{\rm F} = 
\frac{ \left\{ \left\langle  \boldsymbol{\overline{q}\cdot\hat{r}} \right\rangle \right\} }
{ \left\{ \left\langle - k \partial\overline{T}/\partial r \right\rangle \right\} } .
\end{equation}
Using (\ref{eq:qdotr=dTcdr}) allows us to recast $\Nus_{\rm F}$ as
\begin{equation}
 \frac{ -\frac{k}{h} \int (\mathrm{d}T_{\rm C}/\mathrm{d}r) \:\mathrm{d}r }
{ -\frac{k}{h} \int (\mathrm{d}\langle\overline{T}\rangle /\mathrm{d}r) \:\mathrm{d}r } = \frac{ \Delta T_{\rm C} }{ \Delta \langle\overline{T}\rangle } 
= \Nus_{\rm T}.
\label{eq:NuDef}
\end{equation}	
The order of the horizontal and temporal averaging in $\Delta \langle\overline{T}\rangle $ is interchangeable and in practice
we determine $\overline{ \Delta \langle T \rangle}$. 
Note that when considering the non-dimensional form of (\ref{eq:NuDef}) our geometry 
and choice of temperature scaling results in $\Delta T^\star_{\rm C} \approx 1.2$ 
rather than unity, as would be typical for Cartesian geometry \citep[e.g.][]{Otero:2002by, Goluskin:2015iw}
 (see also appendix~\ref{app:DeltaTC}).

Let us consider the implications of a change in Nusselt number between two models with the same value of $\Ray_{\rm F}$
but different patterns or amplitudes of boundary heat flux variation.
Equation~(\ref{eq:NuDef}) shows that a change in $\Nus$ corresponds to a change in the temperature drop across the system
since $\Delta T_{\rm C}$ is set by $\beta$ (recall equation~\ref{eq:DeltaTCondShell}), and hence by $\Ray_{\rm F}$. 
Furthermore, equations~(\ref{eq:qtotaldef}) and (\ref{eq:qdotr=dTcdr}) imply that any change in $\langle \partial\overline{T}/\partial r\rangle$ 
must be compensated by a complementary change in $\langle\overline{u_r T}\rangle$.
Changing the efficiency of heat transport requires alteration of both the temperature and velocity fields.

The time-averaged temperature drop across the system can be expressed using the spherical harmonic expansion of the
temperature field as
\begin{equation}
\Delta \langle\overline{T}\rangle = \overline{\vartheta_0^0(\ri)} - \overline{\vartheta_0^0(\ro)};
\end{equation}
any change in $\Nus_{\rm T}$ induced by the heterogeneous boundary condition must influence the $Y_0^0$ spherical harmonic 
component of the temperature field. The imposed patterns of boundary heterogeneity have zero mean and hence
the homogeneous and heterogeneous thermal boundary conditions have identical $Y_0^0$ components for a given $\Ray_{\rm F}$. 
Since there is no interaction between harmonics in the diffusive part of the energy equation~(\ref{eq:EnergyTDim}), any change in
$\overline{\Delta T}$, and hence $\vartheta_0^0(r)$, must arise from non-linear interaction between the flow and temperature fields. 
This can be seen by writing  equation~(\ref{eq:qdotr=dTcdr}) as
\begin{equation}
\overline{\langle u_rT \rangle} = \kappa \frac{d}{dr}\left(\overline{ \vartheta_0^0} - T_{\rm C}\right).
\label{eq:urTalternate}
\end{equation}
The radial component of velocity depends only on the poloidal component of the velocity field, 
thus spherical harmonic expansion of the variables on the left-hand side of (\ref{eq:urTalternate}) leads to
\begin{equation}
\frac{1}{r} \sum_{\ell,m} \left(\frac{\ell(\ell+1)}{2\ell+1}\right) \overline{ p_\ell^m \vartheta_\ell^m}
= \kappa \frac{d}{dr}\left(\overline{ \vartheta_0^0} - T_{\rm C}\right),
\label{eq:urTequivalence}
\end{equation}
where we have made use of the orthogonality of the (Schmidt-normalised) spherical harmonics. Equation~(\ref{eq:MassDim}) implies that $p_0^0 = 0$ and
thus we see that $\vartheta_0^0$ must be modified by interactions between flow and temperature at other harmonics. 
 
Consider, for example, a homogeneous model to which is added a $Y_1^1$ heat flux heterogeneity at the outer boundary. 
In this case we anticipate that the heat flux heterogeneity will increase 
$\overline{\vartheta_1^1}$ near the top of the fluid core, relative to the homogeneous boundary case. 
This temperature perturbation in the core then generates, on average, some amount of $p_1^1$ flow by promoting (inhibiting) 
downwelling in regions of enhanced (reduced) outward heat flux across the boundary. 
Any resultant increased correlation between the hemispheric patterns in radial flow and temperature (i.e., an increase
in $\overline{p_1^1 \vartheta_1^1}$ for radii near $\ro$) implies an altered $d\overline{\vartheta_0^0}/dr$
 and hence $\Nus_{\rm T}$. 
Correlations at harmonics other than those of the imposed heterogeneity could also be promoted
by the boundary heterogeneity through some more complex set of dynamics. 
Regardless, for fixed $\Ray_{\rm F}$ the total heat transport through the system remains unchanged and
an increase in $\Nus$  with heterogeneous boundary conditions, 
which reflects a repartitioning of heat transport from conduction to advection, requires
an increased correlation between $u_r$ and $T$ and a smaller average radial temperature gradient. 
This reorganisation of flow and temperature fields could occur throughout the domain or be limited to a relatively restricted region, 
for example near the outer boundary. 
In the following section we present the heat transport results for our suite of models. 
We will not present a detailed analysis of the association between flow and $\Nus$ for all simulations,
but any case for which the heterogeneous boundary conditions have altered $\Nus$ relative to the equivalent homogeneous 
case must have some reorganisation of the time-averaged flow in accordance with the principles outlined here.

\section{Results and discussion}\label{sec:results}

\subsection{Numerical model, parameters, and convergence tests}\label{sec:ParamsAndConverge}

The pseudo-spectral method used in this work is described in \citet{Willis:2007cu}; it passes the dynamo benchmark and performs comparably to other 
pseudo-spectral methods \citep{Matsui:2016hg}. The velocity field is decomposed into toroidal and poloidal scalars, which ensures that the divergence-free 
condition is satisfied exactly. All scalars are then expanded in Schmidt-normalised spherical harmonics on each spherical surface and represented in radius 
by second-order finite differences. The finite difference points are located at the zeros of the Chebyschev polynomials, providing finer spacing near the 
upper and lower boundaries. Timestepping is accomplished in spectral space using a predictor-corrector scheme that treats diffusion terms implicitly, 
while the Coriolis, buoyancy and nonlinear terms are treated explicitly. Nonlinear terms are transformed into real space at each timestep using the 
spherical transform method \citep{Orszag:1971if}. At each radius multiplications are performed on a Gauss-Legendre grid with $\frac{3}{2}{\ell}_{\rm max}$ 
colatitude points and $3{\ell}_{\rm max}$ longitude points. The number of radial grid points, $N_r$, and the 
maximum spherical harmonic degree and order, $\ell_{\rm max} = m_{\rm max}$, for all runs are given in appendix~\ref{app:Tables}.
Our choices of $\ell_{\rm max}$ are similar to those used by \citet{Gastine:2016gq} for comparable control parameters.

In this work we focus on the global heat transport of rotating convection in a spherical shell with variable heat flux boundary conditions.
To do so we have run a suite of 106 numerical simulations with: Ekman number, $\Ek = 10^{-4}$, $10^{-5}$, $10^{-6}$; flux Rayleigh number, 
$3 \times 10^5 \leq \Ray_{\rm F} \leq 1.8 \times 10^{10}$; Prandtl number, $\Pran =1$. Strong rotation inhibits the onset of convection; 
for our thermal and mechanical boundary conditions, and choices of 
Ekman and Prandtl numbers, linear stability analysis of the homogeneous cases
(for details see \citet{Gibbons:2007ef} and \citet{Davies:2009il}) indicates that our simulations fall in the range 
$1.2 \Ray_{\rm C}  \lesssim \Ray_{\rm F} \lesssim 800 \Ray_{\rm C}$ (the critical Rayleigh number, $\widetilde{\Ray}_{\rm C}$, 
and most unstable mode at onset, $m_{\rm C}$, for our cases are given in table~\ref{tab:Critical}). 
The control and output parameters for all runs are detailed in appendix~\ref{app:Tables}.

\begin{table}
  \begin{center}
\def~{\hphantom{0}}
  \begin{tabular}{ccc}
      $\Ek$  & $\widetilde{\Ray}_{\rm C}$  &  $m_{\rm C}$ \\[3pt]
       $10^{-4}$  & 16.4 & 5 \\
       $10^{-5}$  & 24.7 & 12 \\ 
       $10^{-6}$  & 41.0 & 25 \\ \end{tabular}
  \caption{Critical Rayleigh number and critical azimuthal wavenumber for our simulations.}
  \label{tab:Critical}
  \end{center}
\end{table}

 We consider three different patterns of heat flux imposed at the outer boundary.
Simulations with a homogeneous outer boundary have $q_{\rm L}^\star=0$. 
Cases with the hemispheric pattern described by the $Y_1^1$ spherical harmonic (left panel, figure~\ref{fig:Patterns}) 
are referred to using ${\rm H}q_{\rm L}^\star$, in these
simulations $q_{\rm max}$ ($q_{\rm min}$) is aligned with the negative (positive) $x$ axis. Cases with boundary 
heterogeneity derived from the observed pattern of seismic velocity variations in the lowermost mantle
 \citep{Masters:1996ey} (right panel, figure~\ref{fig:Patterns}) are referred to using ${\rm T}q_{\rm L}^\star$. 
 The amplitude of the heat flux heterogeneity is set to ${\rm H}q_{\rm L}^\star, {\rm T}q_{\rm L}^\star=2.3$ or 5.0; values based on  
a proposed scaling from seismic velocity to temperature following the work of \citet{Nakagawa:2008kk}.

\begin{figure}
  \centerline{\includegraphics{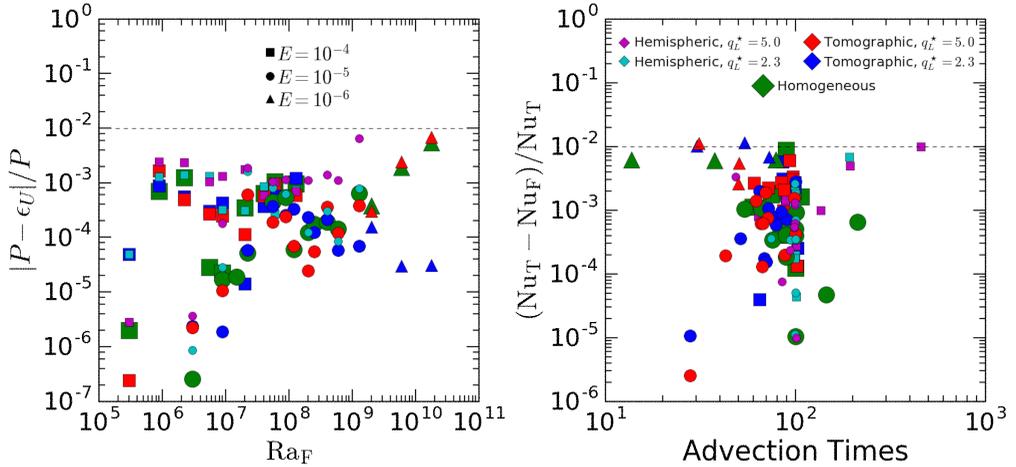}}% Images in 100% size
  \caption{Left panel: Convergence of all models as measured by the difference between the time average of the buoyancy production, $P$,
  and viscous dissiaption, $\epsilon_U$, both integrated over the volume of the shell.
  Right panel: Convergence of the time-averaged Nusselt number for all models as measured by the difference between $\overline{\Nus_{\mathrm F}}$ and 
  $\overline{\Nus_{\mathrm T}}$.
  Symbol shapes indicate the value of $\Ek$, symbol size and colour indicate the nature of the boundary conditions.}
\label{fig:Converge}
\end{figure}

The spatial convergence of each simulation is evaluated by checking that the buoyancy production throughout the volume, $P$, is
matched by the viscous dissipation, $\epsilon_U$, in the time average \citep[see e.g.][]{Gastine:2015cj}. 
The left panel of Figure~\ref{fig:Converge} shows $|P - \epsilon_U| / P$ for all cases, this residual is always less than $10^{-2}$. 
We compute the thickness of the viscous boundary layers at $r = \ri ,\ro$ 
based on the location of the local maxima in the radial profile of horizontal velocity variations following the method described in \citet{King:2013fx}) and report
the number of grid points within the boundary layers for each simulation in appendix~\ref{app:Tables}.

After removal of the initial transient, time averages are constructed over a span of at least 10 advection times and in general 
of around 100 advection times. The Reynolds number allows conversion from advection to diffusion times; the durations of
the runs mostly lie between 0.01 and 10 diffusion times. 
Convergence of the Nusselt number is tested by considering the difference between 
$\overline{\Nus_{\mathrm F}}$ and $\overline{\Nus_{\mathrm T}}$ as determined by time averaging over the  run. 
As shown in the right panel of figure~\ref{fig:Converge} the difference between these two methods of 
calculating the Nusselt number is on the order of 1\% or less.

\begin{figure}
  \centerline{\includegraphics{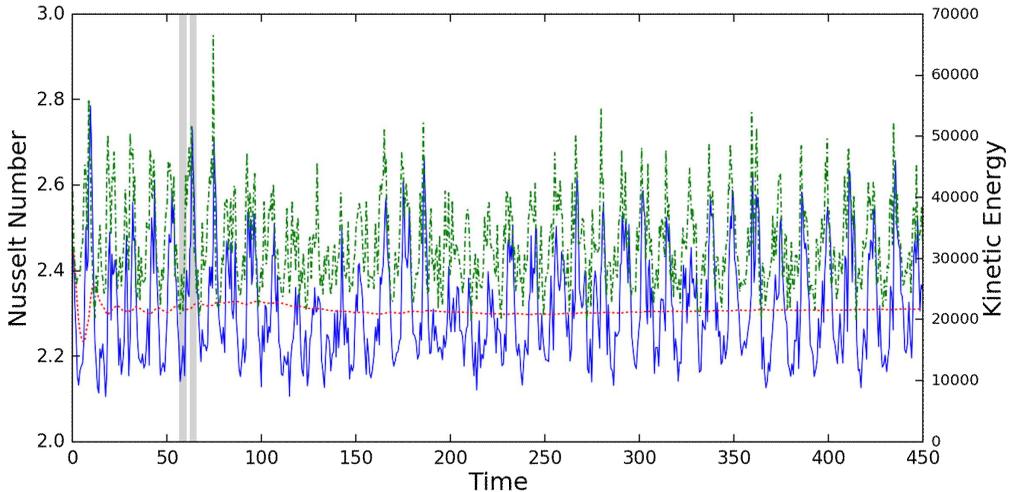}}% Images in 100% size
  \caption{${\rm KE}^\star$ (dash-dot green line, righthand axis), $\Nus_{\rm T}$ (blue solid line, lefthand axis), and the
  running average of $\Nus_{\rm T}$ (dotted red line, lefthand axis) for the run with $\Ek = 10^{-4}$, $\Ray_{\rm F} = 2.25\times 10^6$, and
${\rm H}q_{\rm L}^\star = 5.0$. The representative flow patterns for the low- and high-$\Nus$ states shown in figure~\ref{fig:OscillatingNuFlows}
were found by averaging over the time periods indicated by the grey shading. Time is measured by the advection time scale. }
\label{fig:OscillatingNuCase}
\end{figure}

The case with $\Ek = 10^{-4}$, $\Ray_{\rm F} = 2.25\times 10^6$, and
 ${\rm H}q_{\rm L}^\star = 5.0$ (the right-most point in the right panel of Figure~\ref{fig:Converge}) required
 over 450 advection times to reach our 1\% convergence target, significantly longer than the other runs. 
The kinetic energy and instantaneous $\Nus$ time series
for this case display large amplitude fluctuations (figure~\ref{fig:OscillatingNuCase}). 
The system switches between two states each of which persists for several advection times, 
with times of higher (lower) ${\rm KE}^\star$ correlated with periods of higher (lower) $\Nus$;
$\Rey = 67.3$ for this run, so the period of the oscillations is approximately 0.1 diffusion times.
There is a time lag between the total and zonal kinetic energies of the system 
suggesting that the convection generates progressively stronger zonal flow until the resultant
shear disrupts the convective rolls.
Similar relaxation oscillations have been seen in homogeneous
rotating spheres \citep[see e.g.][for a review]{Busse:2002gv}, here the boundary heterogeneity 
modulates convection activity such that localised convection is concentrated in the hemisphere that
is most strongly cooled by the overlying mantle.

\begin{figure}
  \centerline{\includegraphics{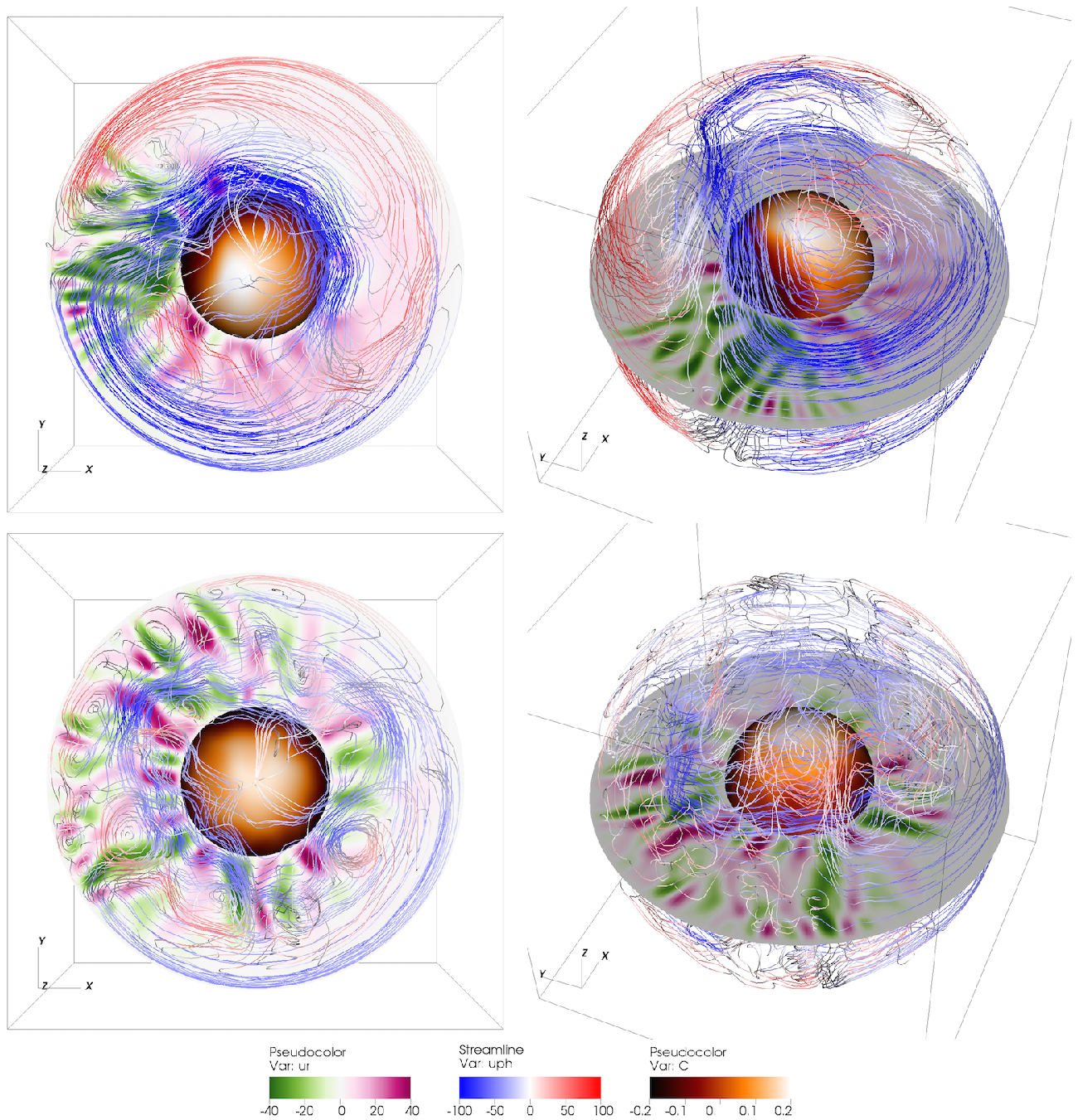}}% Images in 100% size
  \caption{Time average flows for the run with $\Ek = 10^{-4}$, $\Ray_{\rm F} = 2.25\times 10^6$, and
 ${\rm H}q_{\rm L}^\star = 5.0$ averaged over a period of high $\Nus$ (top row) and a period of low $\Nus$ (bottom row),
 the averaging periods are indicated by the grey bands in figure~\ref{fig:OscillatingNuCase}.
The equatorial plane is coloured by $u^\star_r$ in the plane (green-magenta),
streamlines of the time-averaged velocity field are
coloured by $u^\star_\phi$ (blue-red),  and the inner boundary is
coloured by temperature anomaly relative to the horizontal average (brown-orange-white). On the outer boundary $q_{\rm min}$  is
aligned with the positive $x$-axis and $q_{\rm max}$ is aligned with the negative $x$-axis. The rotation vector points in the 
positive $z$ direction.}
\label{fig:OscillatingNuFlows}
\end{figure}

Representative flow patterns for the low- and high-$\Nus$ states were found by averaging over the time periods
indicated by the grey shading in figure~\ref{fig:OscillatingNuCase}, each 
of which corresponds to approximately three advection times.
The flow pattern in the low-$\Nus$ state (lower row of figure~\ref{fig:OscillatingNuFlows}) 
consists of a well-developed zonal flow in the shell interior interacting with convective rolls aligned
with the rotation axis. Relatively hot fluid accumulates near the outer boundary in the positive $x$
hemisphere below $q_{\rm min}$, which tends to suppress the formation of convective rolls, and hence radial flow, above mid-depth in that hemisphere. 
Conversely, $q_{\rm max}$ in the negative $x$ hemisphere tends to enhance radial flow. 
The high-$\Nus$ state (upper row of figure~\ref{fig:OscillatingNuFlows})
is characterised by two large-scale circulations anchored by a large downwelling beneath $q_{\rm max}$.
Peak upwelling velocities in the high-$\Nus$ state  are similar to those of the low-$\Nus$ state; however, peak downwelling
velocities have approximately 50\% greater amplitude. 
The strong downwelling is relatively effective at transporting cold material deep within the shell
such that the high-$\Nus$ state has a long-wavelength azimuthal temperature anomaly at the equator of the inner boundary, 
in addition to the general pattern of positive (negative) temperature anomalies at high (low) latitudes seen at all times in this model. 
The higher average velocities and changed pattern of convection increase the global correlation between radial velocity and 
temperature and correspondingly reduce the average temperature drop across the shell during the 
high-${\rm KE}^\star$ -- high-$\Nus$ state.

\subsection{Nusselt-Rayleigh scaling}\label{sec:NuRa}

\begin{figure}
  \centerline{\includegraphics{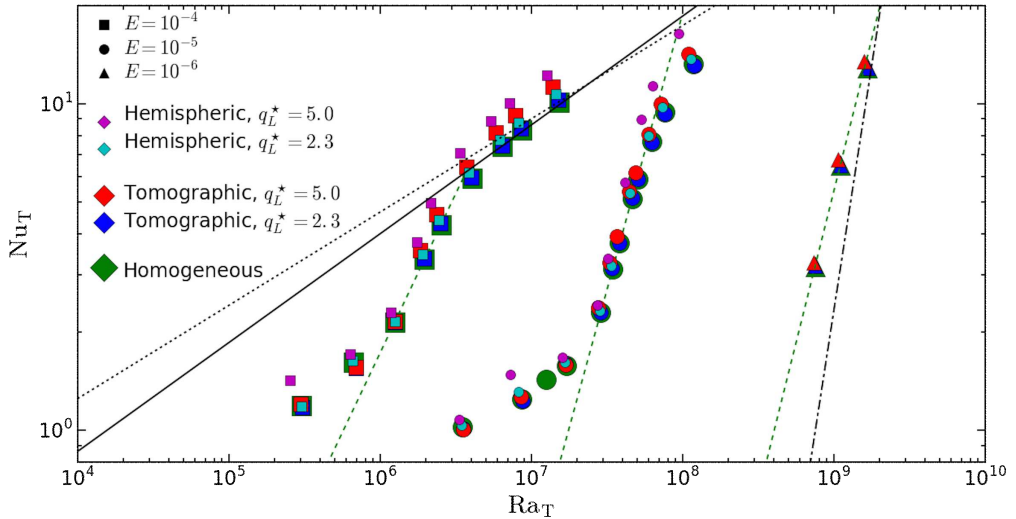}}% Images in 100% size
  \caption{Scaling of $\Nus_{\rm T}$ against $\Ray_{\rm T}$.  Dashed green lines are fits to the $\Ray$ homogeneous cases, giving
  $\Nus \propto \Ray_{\rm T}^{0.98}$,  $\Nus \propto \Ray_{\rm T}^{1.69}$, and $\Nus \propto \Ray_{\rm T}^{1.86}$
   for $\Ek = 10^{-4}$, $\Ek = 10^{-5}$, and  $\Ek = 10^{-6}$, respectively. For comparison, black lines are scalings 
   of $\Nus \propto \Ray_{\rm T}^{1/3}$ (solid) and $\Nus \propto \Ray_{\rm T}^{2/7}$ (dotted) for non-rotating convection,   
   and $\Nus \propto \Ray_{\rm T}^3 \Ek^4$ for rotationally constrained convection (dash-dot, for $\Ek = 10^{-6}$).
  Symbol shapes indicate the value of $\Ek$, symbol size and colour indicate the nature of the boundary conditions. }
\label{fig:NuRa}
\end{figure}

\begin{figure}
  \centerline{\includegraphics{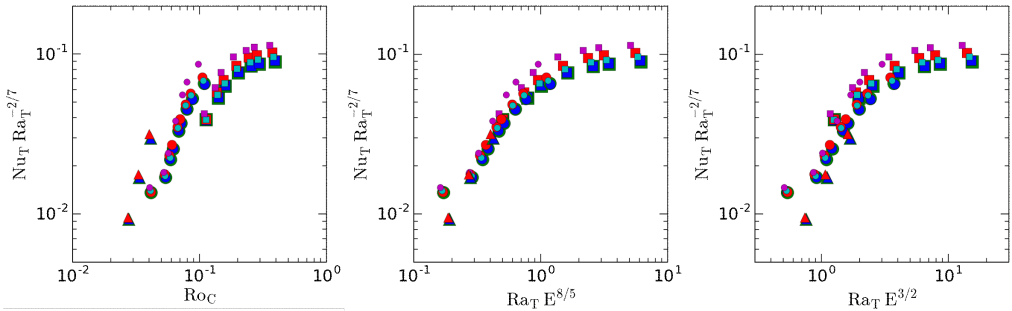}}% Images in 100% size
  \caption{Compensated Nusselt number, $\Nus \Ray^{-2/7}$, against proposed parameters controlling the transition 
  out of the rotationally dominated regime. For clarity, and to focus on behaviour above the weakly non-linear regime, 
  only cases with $\Ray_{\rm F} > 7\Ray_{\rm C}$ are shown.  
  Left panel: $\Ros_{\rm C} = (\Ray \Ek^2 / \Pran)^{1/2}$, the convective Rossby number  \protect\citep{Gilman:1977fj}.
  Middle panel: $\Ray \Ek^{8/5}$, based on the local Rossby number of the thermal boundary layer \protect\citep{Gastine:2016gq}.
  Right panel:  $\Ray \Ek^{3/2}$, based on thermal and viscous boundary layer crossing \protect\citep{King:2012gi}.
  Symbol shapes indicate the value of $\Ek$, symbol size and colour indicate the nature of the boundary conditions, as in figure~\ref{fig:Converge}. }
\label{fig:NuTransition}
\end{figure}

Figure~\ref{fig:NuRa} plots $\Nus_{\rm T}$ against $\Ray_{\rm T}$ for all of our simulations.  For a given value of $\Ek$ the slope of the
 $\Nus$-$\Ray_{\rm T}$ scaling is shallow at relatively low $\Ray$ in the weakly non-linear regime,
steepens as the Rayleigh number increases, and
shallows again at the highest values of $\Ray_{\rm T}$, particularly for the runs with $\Ek = 10^{-4}$ 
(see also figures~\ref{fig:NuTransition} and \ref{fig:NuRaComp} for plots of several compensated $\Nus$ scalings).
Due to the computational expense we have performed a limited number of runs at $\Ek = 10^{-6}$, concentrating
on $\Ray$ values where we expect to be in a regime of non-linear rotating convection, which is our main region of interest.
\citet{Gastine:2016gq} produced a regime diagram (their figure 20) showing that the rapidly-rotating regime is expected for only
a small span of $\Ray_{\rm T}$ for our values of $\Ek$, being bounded below by the weakly non-linear regime (when $\Ray \leq 6 \Ray_{\rm C}$) 
and above by a regime they term transitional, in which rotational effects no longer dominate even if the effectively non-rotating regime has not 
been reached. Our model has a different aspect ratio, radial gravity profile, and thermal boundary conditions than \citet{Gastine:2016gq} so
an exact correspondence between our results and their regime diagram is not to be expected; however, we observe similar regime transitions.
  
There is a clear decrease in the slope of the $\Nus \propto \Ray_{\rm T}^{\gamma}$ scaling for the 
highest $\Ray_{\rm T}$ cases with $\Ek = 10^{-4}$; we do not, however, have
sufficient results at high Rayleigh to adequately determine a best-fit scaling. As examples, in figure~\ref{fig:NuRa}
we plot both $\Nus \propto \Ray_{\rm T}^{1/3}$ and $\Nus \propto \Ray_{\rm T}^{2/7}$ scalings
for comparison with the highest Rayleigh number results with $\Ek = 10^{-4}$. It is unlikely that a single scaling
is appropriate over a wide range of Rayleigh number. Previous work in spherical geometry found that continuous changes in flow properties 
occur within the transitional regime (where our high-$\Ray_{\rm T}$,
$\Ek = 10^{-4}$ results lie), including a reduction in the exponent of the  $\Nus$-$\Ray_{\rm T}$ scaling as supercriticality 
increases and the importance of rotational forces is progressively reduced \citep{Gastine:2015cj, Gastine:2016gq}.

The transition from rotationally-dominated to non-rotating convection corresponds to buoyancy becoming dominant over Coriolis
forces; the most appropriate parameterisation of this transition remains an open question. 
In figure~\ref{fig:NuTransition} we plot $\Nus \Ray_{\rm T}^{-2/7}$ against three proposed transition parameters;
to focus on the transition to the non-rotating regime at high Rayleigh number we plot only runs with $\Ray_{\rm F} > 7\Ray_{\rm C}$, which
removes cases within the weakly nonlinear regime identified by \citet{Gastine:2016gq}.
The global-scale force balance can be
expressed by the convective Rossby number, $\Ros_{\rm C} = ( \Ray_{\rm T} \Ek^2 / \Pran )^{1/2}$, such that
the transition to non-rotating convection might be expected when $\Ros_{\rm C} = {\it O}(1)$ \citep{Gilman:1977fj, Zhong:2009cx}. 
We find that all of our runs have $\Ros_{\rm C} < 1$ (left panel of figure~\ref{fig:NuTransition}) 
and do not support a $\Ros_{\rm C} = {\it O}(1)$ transition parameter, which agrees with previous 
numerical and laboratory studies \citep{King:2009hy, Stellmach:2014kp, Cheng:2015ka}.
Given the importance of the Ekman and thermal boundary layers in controlling heat transport through the system \citet{King:2012gi}
used the intersection of scaling laws for 
the thicknesses of these two boundary layers to suggest that the transition should occur for $\Ray_{T} E^{3/2} = {\it O}(1)$.
If the transition is governed by the loss of geostrophic balance within the thermal boundary layer, scaling of
the local Rossby number for the layer leads to a transition occurring at $\Ray_{T} E^{8/5} = {\it O}(1)$ \citep{Julien:2012dc, Gastine:2016gq}.
This latter scaling does a somewhat better job of collapsing our results 
(compare the middle and right panels of figure~\ref{fig:NuTransition}).
We would require significantly more high-$\Ray$ runs to properly characterise both the slope of 
the $\Nus$-$\Ray_{\rm T}$ scaling and the transition parameter 
 for the regime in which Coriolis effects no longer dominate buoyancy. Figures~\ref{fig:NuRa} and 
 \ref{fig:NuTransition} suggests that few (if any) of our
 $\Ek = 10^{-4}$ runs fall within the non-linear and rotationally-dominated regime, but our runs at lower Ekman
 do sample this regime, a result consistent with the regime diagram of \citet{Gastine:2016gq}.

We fit straight lines to each set of four consecutive runs (in terms of their $\Ray$) for the $q_{\rm L}^\star =0$ simulations at 
$\Ek = 10^{-4}$ and $10^{-5}$ and take the line of best fit with maximum slope as the $\Nus$-$\Ray_{\rm T}$ scaling for the
rotating regime. Although all of the runs that end up contributing to this fit may not be rotationally dominated, 
they are at least rotationally influenced. 
In both cases the line of steepest slope also corresponds to the four consecutive runs that are best fit by a straight line,
although we note that these fits span a limited range of $\Ray_{\rm T}$ values.
For $\Ek = 10^{-6}$ we fit a straight line to the three  $q_{\rm L}^\star =0$ simulations. 
As the Ekman number decreases, the exponent of the $\Nus$-$\Ray_{\rm T}$ scaling 
increases from $0.98$ for $\Ek = 10^{-4}$, to $1.69$ for $\Ek = 10^{-5}$,
to $1.86$ for $\Ek = 10^{-6}$; such steepening of the $\Nus$-$\Ray_{\rm T}$ scaling
with decreasing $\Ek$ has previously been seen in both numerical and laboratory studies of rotating convection in a variety of geometries
\citep[e.g.][]{King:2012gi, Cheng:2015ka, Gastine:2016gq}. 
Our scaling is steeper than the 
diffusivity-free scaling of $\Nus \propto \Ray_{\rm T}^{3/2} \Ek^2$ expected at low $\Ek$ for convection with free-slip boundaries
\citep{Gillet:2006be, Julien:2012dc, Stellmach:2014kp}.
We employ no-slip boundary conditions for which the effect of Ekman pumping has been shown 
to increase the efficiency of heat transport and hence the slope of the $\Nus$-$\Ray_{\rm T}$ scaling 
\citep{Stellmach:2014kp, Aurnou:2015jh, Julien:2016js}.
Although the scaling exponents for our low-$\Ek$ runs lie above the diffusivity-free scaling, they are well below
the scaling exponents of $\sim 3$ found in studies with no-slip boundaries in cylindrical and Cartesian geometries
at similar $\Ek$ \citep{King:2012gi, Cheng:2015ka}.

\subsection{$\Nus$ enhancement by heterogeneous boundaries}\label{sec:NuEnhance}

\begin{figure}
  \centerline{\includegraphics{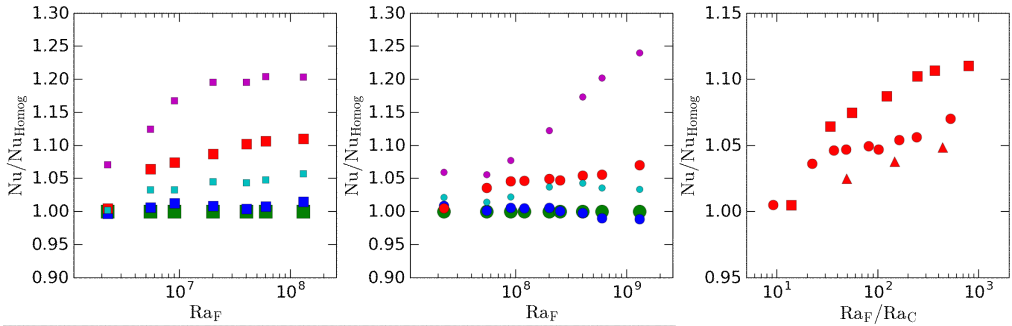}}% Images in 100% size
  \caption{Compensated versions of the Nusselt number as a function of Rayleigh number. In each case
  $\Nus_{\rm T}$ is divided by the value of $\Nus_{\rm T}$ for the homogeneous case at equivalent $\Ray_{\rm F}$
  For clarity, and to focus on behaviour above the weakly non-linear regime,
  only cases with $\Ray_{\rm F} > 7\Ray_{\rm C}$ are shown. 
  Symbol colours, shapes, and sizes as in figure~\ref{fig:NuRa}. 
 Left panel: runs with $\Ek = 10^{-4}$. Middle panel: runs with $\Ek = 10^{-5}$
 Right panel: runs with $Tq^\star_{\rm L} = 5.0$ plotted as a function of supercriticallity.  
  }
  \label{fig:NuRaComp}
\end{figure}

Our particular interest is whether different boundary heterogeneities alter the amplitude
of $\Nus$ for a given value of $\Ray$ or even the slope of the associated scaling law;
to more clearly show such differences figure~\ref{fig:NuRaComp} presents compensated $\Nus$ values.
In this figure, the $\Nus$ for each run 
is divided by the value obtained from the $q_{\rm L}^\star =0$ case at the same $\Ray_{\rm F}$.
There is little difference between the ${\rm T}q_{\rm L}^\star = 2.3$ cases and the equivalent $q_{\rm L}^\star =0$ 
cases. However, in the other cases the heterogeneity in outer boundary heat flux tends to enhance the efficiency
of heat transport; the enhancement tends to become greater as $\Ray$ becomes more supercritical, the wavelength of the imposed
boundary heterogeneity increases, or the amplitude of $q_{\rm L}^\star$ increases. 
Investigations of relatively low-$\Ray$ convection in high-$\Pran$ fluids
\citep{Zhang:1993ts, Davies:2009il} have found that the effect of outer boundary heterogeneity on the underlying fluid is greater
when larger wavelength lateral variations are applied, consistent with our results at $\Pran = 1$  and more highly supercritical $\Ray$.

We find that addition of outer boundary heterogeneity tends to increase $\Nus$ relative to equivalent $q_{\rm L}^\star = 0$ cases, 
whereas \citet{Dietrich:2016ba} found a nearly linear reduction in $\Nus$ with increasing ${\rm H}q_{\rm L}^\star$ for simulations with 
$\Ray / \Ray_{\rm C} = 25$, $\Ek = 10^{-4}$ and a range of $0 \leq {\rm H}q_{\rm L}^\star \leq 2.0$ 
(see figures 11 and 13 of their paper and note that their definition of $q^\star$ is equal to one half our ${\rm H}q_{\rm L}^\star$).
This difference arises from the fact that we use  $\overline{\langle \Delta T \rangle}$ in determining $\Nus_{\rm T}$, whereas 
\citet{Dietrich:2016ba} estimate $\Delta T = T_{\rm max} - T_{\rm min}$, where $T_{\rm max}$ and $T_{\rm min}$ are the maximum 
and minimum values of $T$ in the domain (Dietrich, personal communication). When there are large lateral variations in boundary heat 
flux, and hence  fluid temperature, the point-wise maximum approach can significantly overestimate the average temperature drop across 
the system and thus underestimate $\Nus_{\rm T}$. We have found that calculating $\overline{\langle \Delta T \rangle}$ for their simulations 
yields an increase in $\Nus$ for boundary-forced cases compared to the corresponding homogeneous case.  

Differences in the Nusselt number between our ${\rm H}q_{\rm L}^\star = 5.0$ and  
$q_{\rm L}^\star =0$  cases at equivalent $\Ray_{\rm F}$ can be as much as 20--25\% and appear to saturate as the simulations reach the regime where
rotation no longer dominates the force balance; this saturation effect is most evident for our $\Ek = 10^{-4}$ cases (left panel of
figure~\ref{fig:NuRaComp}) as the regime without rotational dominance is more easily reached for larger Ekman number.  
 Since the difference in $\Nus$ between the  ${\rm H}q_{\rm L}^\star = 5.0$ and  $q_{\rm L}^\star =0$  cases  grows with $\Ray$ in
the rotationally-dominated regime, they are characterised by different exponents for the $\Nus$-$\Ray_{\rm T}$ scaling; for example,
we find an exponent of 2.05 for $\Ek = 10^{-5}$ and ${\rm H}q_{\rm L}^\star = 5.0$, significantly above the 1.69 exponent for $q_{\rm L}^\star = 0$.
Although each scaling determination uses only four simulations from a relatively limited
range of $\Ray_{\rm T}$ the enhanced efficiency of heat transport in the ${\rm H}q_{\rm L}^\star=5.0$ cases is clear and
much greater than the ${\rm H}q_{\rm L}^\star=2.3$ cases, which reach at most $\sim5$\% enhancement.
For the ${\rm T}q_{\rm L}^\star = 5.0$ cases enhancements of about 5--10\% are obtained for the three Ekman numbers we consider, with
a decrease in the  enhancement of $\Nus$  as $\Ek$ is lowered 
(right panel of figure~\ref{fig:NuRaComp}). For our lowest $\Ek$ cases we have not been able to reach the regime where rotation 
no longer dominates the force balance; we expect that the enhancement of the Nusselt number would continue to
increase with supercriticality within the rapidly-rotating regime, before saturating at sufficiently large $\Ray_{\rm F}/\Ray_{\rm C}$.

\begin{figure}
  \centerline{\includegraphics{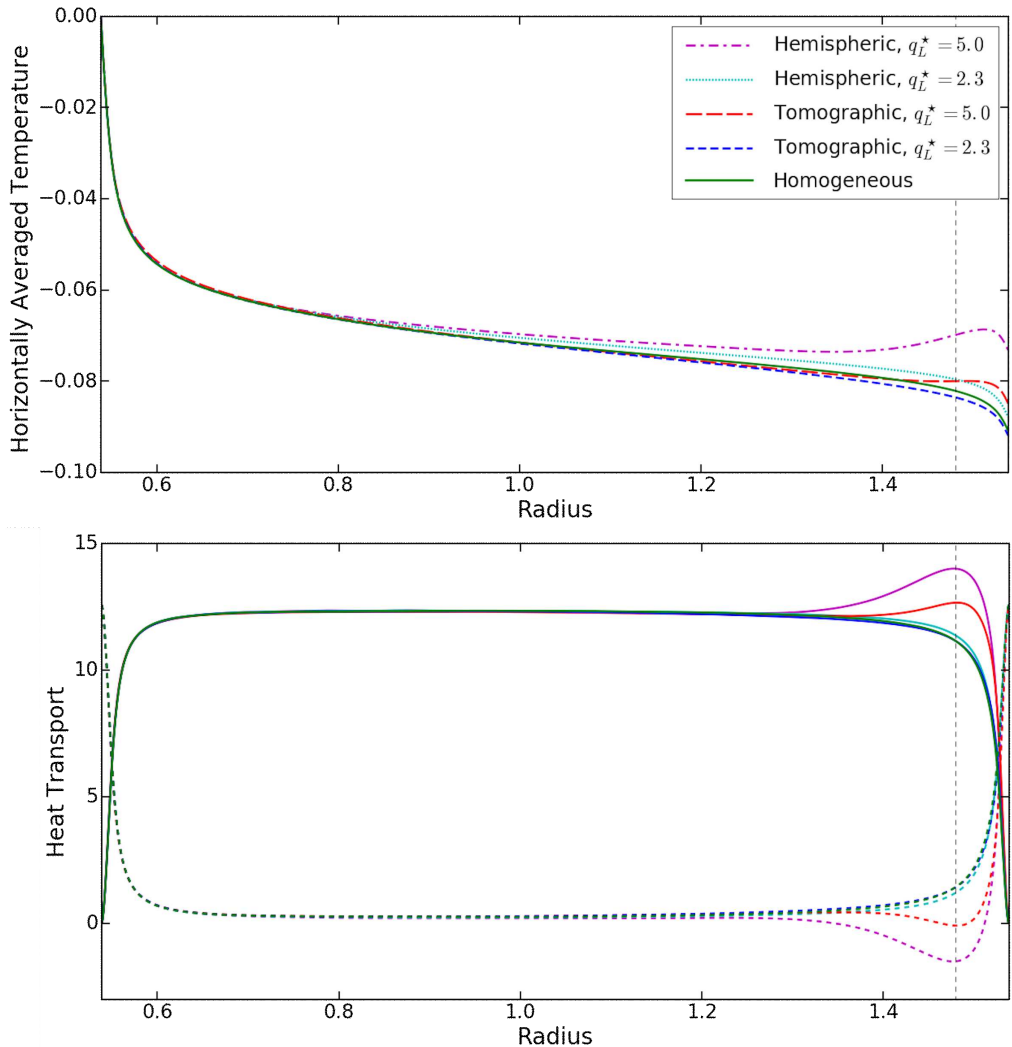}}% Images in 100% size
  \caption{Top panel: radial profiles of $\langle\overline{T^\star}\rangle$ for runs with $\Ek = 10^{-5}$ and $\Ray_{\rm F} = 1.3\times 10^9$;
  boundary conditions are indicated by colour and line style. In all cases temperature is set to zero on the inner boundary for the purpose of plotting.
  Bottom panel:  profiles of temporally averaged radial heat flow for these runs,
  solid lines indicate the advective contribution, dashed lines indicate the diffusive contribution;
  the boundary conditions are indicated by colour as in the top panel.  The vertical dashed line indicates the radius for which
  maps and spectra are plotted in figure~\ref{fig:MapsAndSpectra}.  }
  \label{fig:Profiles}
\end{figure}

To better understand the $\Nus$ enhancement by boundary heterogeneity we consider, as an example,
the simulations with $\Ek = 10^{-5}$ and the highest applied Rayleigh number ($\Ray_{\rm F} = 1.3\times 10^9$).
 In figure~\ref{fig:Profiles} we plot radial profiles of temperature, $\langle\overline{T^\star}\rangle$, 
advective heat transport, $4\pi {r^\star}^2 \langle\overline{u_r^\star T^\star}\rangle$, and diffusive heat transport,
$4\pi {r^\star}^2 \langle\overline{-\partial T^\star/\partial r^\star}\rangle$. 
All five cases have thermal boundary layers at the top and bottom of the shell, with the inner boundary layer being
more pronounced, and a small but non-zero temperature gradient in the shell interior. 
For the purposes of plotting $\langle\overline{T^\star}\rangle$ has been set to zero at the inner boundary; differences
 in $\langle\overline{\Delta T^\star}\rangle$ are thus reflected by the average temperatures plotted at $r_{\rm o}^\star$.
The shape of the temperature profiles for the two $q_{\rm L}^\star = 5.0$ runs is noticeably different near the top of the shell, 
with the development of a local maximum in  
$\langle\overline{T^\star}\rangle$ and hence a region where the radial temperature gradient is positive. There is
a corresponding change in the diffusive contribution to the heat transport such that for 
both of these $q_{\rm L}^\star = 5.0$ runs there is a depth range
near the outer boundary where the net diffusive transport of heat is negative, that is, radially inwards.
Since $4\pi r^2 \langle \boldsymbol{\overline{q}\cdot\hat{r}} \rangle$ is conserved, any change in the
diffusive contribution is offset by an equal but opposite change in the advective contribution.
For the heterogeneous runs shown in figure~\ref{fig:Profiles} 
the modification in heat transport relative to the homogeneous case
 is constrained to the outer regions of the shell, with larger amplitude and longer wavelength heterogeneity 
 resulting in changes that extend more deeply.

\begin{figure}
  \centerline{\includegraphics{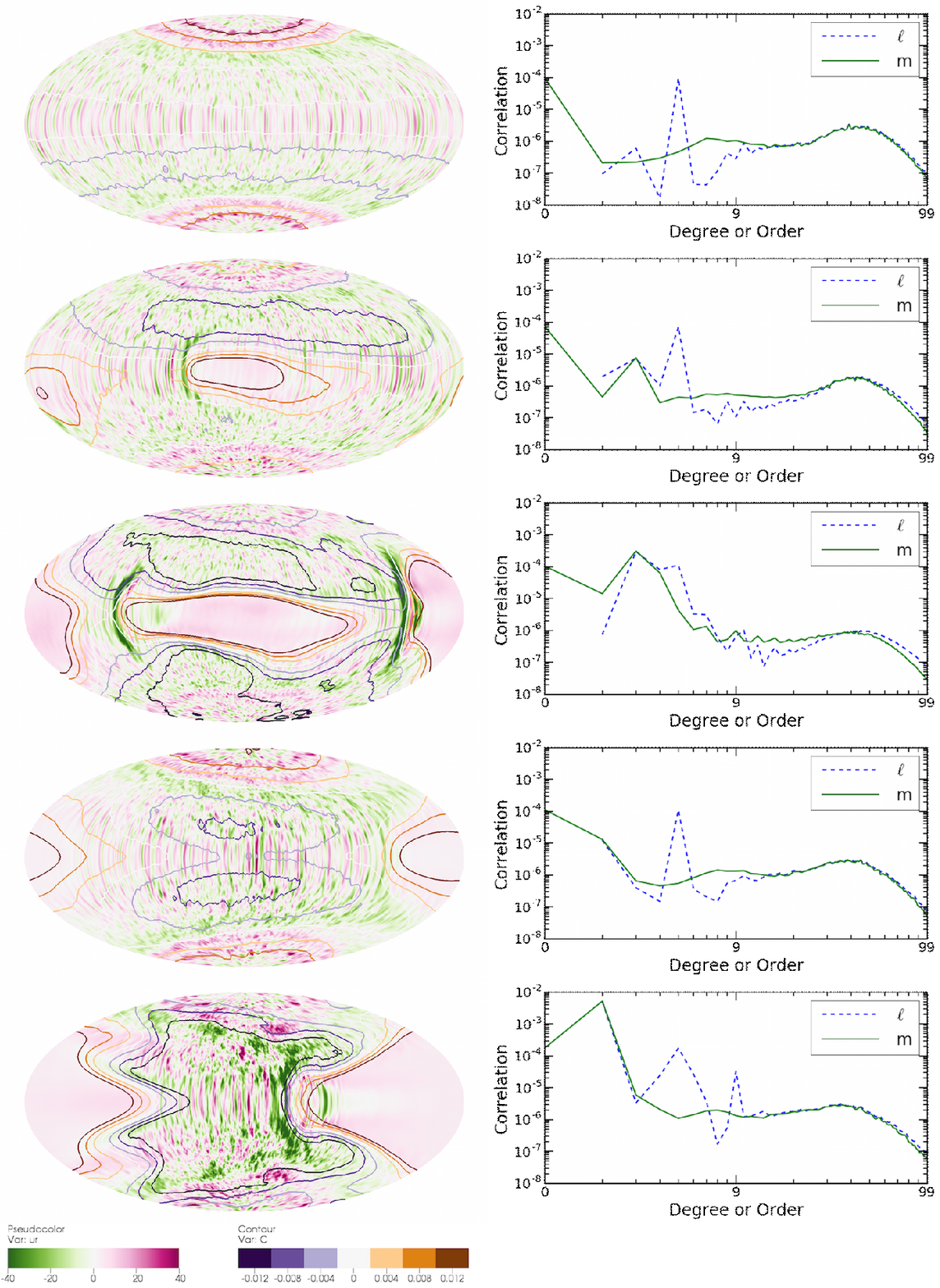}}% Images in 100% size
  \caption{Flow and temperature correlations at
  a radius of $r^\star = 1.48$ for all of the runs with $\Ek = 10^{-5}$ and $\Ray_{\rm F} = 1.3\times 10^9$;
  from top to bottom the boundary conditions are: $q_{\rm L}^\star =0$, ${\rm T}q_{\rm L}^\star = 2.3$, 
   ${\rm T}q_{\rm L}^\star = 5.0$,  ${\rm H}q_{\rm L}^\star = 2.3$,  ${\rm H}q_{\rm L}^\star = 5.0$.
  Left: maps of radial velocity (green-magenta) overlain with contours of temperature anomaly (purple-orange);
  the projection is centred on the negative $x$-axis, thus $q_{\rm max}$ is centred for the ${\rm H}q_{\rm L}^\star$ cases
  and one of the $q_{\rm min}$ is approximately centred for the ${\rm T}q_{\rm L}^\star$ cases. 
  Right: spectra of $\overline{u^\star_r T^\star}$; $\ell$-spectra (i.e.\ integration over all $m$ and $r$ for fixed $\ell$, dashed blue line), 
  $m$-spectra (i.e.\ integration over all $\ell$ and $r$ for fixed $m$, solid green line). For the purpose of plotting, spectra are truncated at $\ell = m = 99$.}
  \label{fig:MapsAndSpectra}
\end{figure}

In figure~\ref{fig:MapsAndSpectra} we plot maps of $ \overline{u^\star_r} $ overlain by contours of
temperature anomaly, $\overline{T^\star} - \langle \overline{T^\star} \rangle$, 
as well as the spectra of $ \left(\frac{\ell(\ell+1)}{2\ell+1}\right) \overline{ p_\ell^m \vartheta_\ell^m}$. In all cases these
quantities have been determined for a radius of $r^\star = 1.48$ (the vertical dashed line in figure~\ref{fig:Profiles}), 
which lies near the peak in advective heat transport for the $q_{\rm L}^\star = 5.0$ cases.
For the $q_{\rm L}^\star = 0$ case the largest single contribution to the advective heat transport
comes from the $Y_4^0$ harmonic; the small-scale convection makes a secondary contribution, with a broad peak
in the correlation between $u_r^\star$ and $T^\star$ centred at spherical harmonics of approximately degree and order 40. 
For this $q_{\rm L}^\star = 0$ case the polar regions are relatively hot in the time average
 and temperature anomalies at low to mid-latitudes are weak.
 For the runs with heterogeneous outer boundaries there is an increased correlation between
$u_r^\star$ and $T^\star$ at long wavelengths, with particular enhancement in advective heat transport at spherical harmonics that match
the imposed boundary conditions. The contributions identified in the spectra of the homogeneous case are still present,
but become relatively less important to the total advective transport as $q_{\rm L}^\star$ is increased. 

For these runs anomalously hot material accumulates near the top of the fluid shell under $q_{\rm min}$.
The formation of small-scale convection rolls is suppressed in these hot regions,
 with a broad region of relatively weak outward flow favoured instead. The heterogeneous runs
also have regions where downwelling is promoted; the time average of the $q_{\rm L}^\star = 5.0$ runs have regions of particularly intense
downwelling near the western edges of the anomalously hot regions. 
Individual snapshots of the flow
show the suppression of small-scale convection under  $q_{\rm min}$; however, the
 focusing of downwelling is not obvious, emerging only in the time average.
The promotion of downwelling at the western boundary of hot regions in these simulations is similar to the jet development observed
by \citet{Sumita:1999bg, Sumita:2002ce} at the front between hot and cold regions in their physical experiment. 
However, in our numerical simulations the
formation of a single spiralling front that spans the shell is prevented by
the presence of strong zonal flows of alternating sign in the shell interior. 
Regardless, the formation of broad regions of weak upwelling and focused regions of enhanced downwelling 
tends to increase advective heat transport
near the top of the shell in these heterogeneous cases and hence raise the Nusselt number relative to the homogeneous case.
For the ${\rm T}q_{\rm L}^\star = 2.3$ case there is increased advective transport relative to
the homogeneous case at some spherical harmonics, most substantially at $Y_2^2$; however, these increases do not offset a reduction in the 
 $Y_4^0$ contribution by $\sim15\%$.
For the ${\rm T}q_{\rm L}^\star = 5.0$ case the reduction in the $Y_4^0$ contribution relative to $q_{\rm L}^\star = 0$ is much smaller
($\sim1\%$) and is more than compensated by increased advective transport at other harmonics.
 For the ${\rm H}q_{\rm L}^\star = 2.3$ case the $Y_4^0$ contribution at this radius is increased relative 
 to the homogeneous case by $\sim10\%$; the relative increase in advective heat transport at $Y_1^1$ is much larger, 
 but the absolute difference is similar for both harmonics.
For the ${\rm H}q_{\rm L}^\star = 5.0$ case the Nusselt number enhancement is dominated by the increased
correlation between radial velocity and temperature at  $Y_1^1$. The results in figure~\ref{fig:MapsAndSpectra} 
 are for one particular radius and a single set of simulations, but are representative
of the changes seen near the top of the fluid layer when heterogeneous boundary conditions are imposed.

All simulations for which the heterogeneous outer boundary condition increases the Nusselt number must do so by increasing
the radial advective transport of heat over some radial extent, which is
generally restricted towards the outer regions of the shell in our simulations.
There is a corresponding reduction in the diffusive contribution to radial heat transport
and thus $\langle \overline{\partial T/\partial r} \rangle$ becomes less negative in the affected region, relative
to the equivalent homogeneous case.
In some of our simulations (for example, the $q_{\rm L}^\star = 5.0$ cases in figure~\ref{fig:Profiles})
the time-averaged temperature gradient even becomes positive in a region near the outer boundary,
an apparent stable stratification. In principle, our use of the Boussinesq
approximation could be invalidated in such cases. The Boussinesq approximation requires that density variation across the system
in the background state is sufficiently small  \citep{Spiegel:1960kt}. Although commonly used in models of planetary cores,
this thin layer approximation is only marginally satisfied for Earth's core \citep[see e.g.][]{Jones:2015bi}, which remains true for our investigation.
The Boussinesq approximation would also be invalid if the system dynamics produce sufficiently large fluctuations in 
density \citep{Spiegel:1960kt}. Although
the time-averaged temperature anomalies in our most extreme heterogeneous cases are large compared to those found in homogeneous 
convection, the fluctuations at any point in time remain smaller than the static variations of the background state and the Boussinesq approximation
still holds.

 The large lateral variations in the time-averaged temperature near the top of the shell that arise in these cases are
 associated with strong variations in local dynamics, with regions of both inhibited and
enhanced small-scale convection (figure~\ref{fig:MapsAndSpectra}). 
This situation is in some sense the complement of the physical experiments of \citet{Alboussiere:2010cy} investigating
stratification at the bottom of the core due to partial melting of the inner core;
they injected fluids that were both compositionally dense and buoyant  
at the bottom of a tank and found that the upwellings of buoyant fluid did not
prevent the formation of a dense, stably stratified layer. 
We have strong lateral variations in thermal buoyancy generated at the top of the shell; however, 
the strong localised convection does not preclude the formation of an average stratification, as in the physical experiment.
Stratification at the top of Earth's core has been hypothesised based on
both seismic and geomagnetic observations, with both thermal and compositional stratification mechanisms proposed
\citep[see e.g.][]{Helffrich:2013bz}.
Heterogeneous outer boundary conditions that increase $\Nus$ by reorganising flow near the top of the core
will necessarily make the average radial temperature gradient more positive in the affected region
and can potentially create a thermal stratification signature in the average temperature profile.

\section{Conclusions}\label{sec:conclusions}

In planetary settings it is expected that long wavelength variations in the heat flux at the top
of metallic cores beneath convecting silicate mantles will be common, although the pattern and amplitude of these variations are
uncertain and will vary both between bodies and through time. 
We have performed 106 numerical simulations, with three Ekman numbers and five different thermal boundary conditions
to investigate how heat transport by thermal convection in rotating spherical shells is impacted by the inclusion of 
 heterogeneous heat flux at the outer boundary.
The  large amplitude and long wavelength boundary heterogeneity we considered tends to increase the Nusselt number
of the system relative to equivalent homogeneous cases (table~\ref{tab:results}).

 The size of the Nusselt number enhancement tends to increase as the amplitude and
wavelength of the boundary heterogeneity increases. The enhancement also tends to increase
 with the supercriticality of the system, although it may saturate as
the system enters the regime in which rotation no longer dominates the force balance. 
This Rayleigh-dependent enhancement can significantly steepen the $\Nus-\Ray_{\rm  T}$ scaling within the rotationally dominated
regime, particularly for the $q_{\rm L}^\star = 5.0$ cases. The Nusselt number enhancement arises from an increased correlation
between radial flow and temperature, particularly near the top of the shell, due to
the development of regions of broad weak upwelling with relatively narrow regions
at their western edge where downwelling is strongly promoted. The fixed-flux boundary conditions require that any increase 
in the advective contribution to the time-averaged radial heat transport is accompanied by a decrease in the diffusive contribution
and hence a modification of the time-averaged temperature profile. In our simulations these effects generally occur near the
top of the shell and in some cases can produce an apparent thermal stratification in the time-average temperature profile, despite
the presence of regions of strong convection at all radii.

\begin{table}
  \begin{center}
\def~{\hphantom{0}}
  \begin{tabular}{cccc}
      $\Ek$  & Boundary  &  $\Nus$-$\Ray_{\rm T}$  & Maximum $\Nus$ \\
                 &  Condition  &   Exponent &  Enhancement  \\[3pt]
       $10^{-4}$  & $q_{L}^\star = 0$                & 0.98 & N/A \\
       $10^{-4}$  & ${\rm T}q_{L}^\star = 2.3$  & 1.07 &  ~1.5\% \\
       $10^{-4}$  & ${\rm T}q_{L}^\star = 5.0$  & 1.25 &  11.1\% \\
       $10^{-4}$  & ${\rm H}q_{L}^\star = 2.3$  & 1.02 &  ~5.8\% \\
       $10^{-4}$  & ${\rm H}q_{L}^\star = 5.0$  & 1.19 &  20.5\% \\
       $10^{-5}$  & $q_{L}^\star = 0$                & 1.69 & N/A \\
       $10^{-5}$  & ${\rm T}q_{L}^\star = 2.3$  & 1.73 &  ~0.9\% \\
       $10^{-5}$  & ${\rm T}q_{L}^\star = 5.0$  & 1.80 &  ~7.0\% \\
       $10^{-5}$  & ${\rm H}q_{L}^\star = 2.3$  & 1.79 &  ~4.9\% \\
       $10^{-5}$  & ${\rm H}q_{L}^\star = 5.0$  & 2.05 &  24.1\% \\ 
              $10^{-6}$  & $q_{L}^\star = 0$                & 1.86 & N/A \\
       $10^{-6}$  & ${\rm T}q_{L}^\star = 2.3$  & 1.85 &  $<0.1$\% \\
       $10^{-6}$  & ${\rm T}q_{L}^\star = 5.0$  & 1.95 &  ~4.9\% \\ \end{tabular}
  \caption{Heat transport enhancement results organised by Ekman number and applied boundary condition:
  exponent of the $\Nus\propto\Ray_{\rm T}^\gamma$ scaling in the rotationally dominated regime; and the largest seen
  enhancement of $\Nus$ relative to the equivalent homogeneous $\Ray_{\rm F}$ case.}
  \label{tab:results}
  \end{center}
\end{table}

\begin{acknowledgements}

CJD is supported by Natural Environment Research Council independent research fellowship NE/L011328/1 and 
a Green scholarship at IGPP.  Wieland Dietrich and Jon Aurnou and thanked for fruitful discussion and three reviewers for their suggestions
that improved this work. This work used the ARCHER UK National Supercomputing Service (http://www.archer.ac.uk)
and ARC2, part of the High Performance Computing facilities at the University of Leeds, UK.
Figures were produced using VisIt \citep{Childs:2012ug} and Matplotlib \citep{Hunter:2007ih}.

\end{acknowledgements}

\newpage

\appendix
\section{The non-dimensionalisation of $\Delta T_{\rm C}$}\label{app:DeltaTC}
The temperature drop across the spherical shell with fixed-flux boundary conditions
in our pure conduction case is $\Delta T_{\rm C} = \beta h / \ri\ro$
 (equation~\ref{eq:DeltaTCondShell}). After non-dimensionalisation of length by $h = \ro - \ri$ and
temperature by $\beta/h$ we have
\begin{equation}
\Delta T^\star_{\rm C}  = \frac{1}{\ri^\star \ro^\star}.
\end{equation}
Therefore, when expressed in terms of the non-dimensional temperature we have
\begin{equation}
\Nus_{\rm T^\star} = \frac{\ri^\star \ro^\star}{\Delta \langle  \overline{T^\star} \rangle }.
\end{equation}
We set our model to match Earth, so $\ri=1.22\times10^6$~m, $\ro=3.48\times10^6$~m and thus 
$\Nus_{\rm T^\star} \approx 1.2/\Delta \langle \overline{T^\star} \rangle $.
We highlight this result in contrast to the plane layer geometry for which 
$\Nus_{\rm T^\star} = 1/\Delta \langle \overline{T^\star} \rangle $ \citep[e.g.][]{Otero:2002by}.

Since our temperature scaling depends on shell geometry, it is possible to construct a spherical shell
for which $\Nus_{\rm T^\star} = 1$ as in the plane layer case. The combination of temperature and 
length non-dimensionalisation would then imply
\begin{eqnarray}
1 &=& \frac{1}{\ri^\star \ro^\star}, \\
1 &=& \ro^\star - \ri^\star.
\end{eqnarray}
The solution is
\begin{eqnarray}
\ri^\star = \frac{\sqrt{5}}{2} - \frac{1}{2} = \frac{1}{\varphi}, \\
\ro^\star = \frac{\sqrt{5}}{2} + \frac{1}{2}  = \varphi,
\end{eqnarray}
giving a radius ratio
${\ri}/{\ro} = 1/\varphi^2 \approx 0.382$,
not far from the value of $\sim$0.351 for Earth.

\section{Tables of results}\label{app:Tables}

Summary tables of the model resolution, control parameters, and selected output parameters for all simulations.
In all cases $\Pran =1$ and the radius ratio $\ri/\ro = 0.351$. $N_r$ is the number of radial points within the fluid shell.
$N_{\delta i}$ and $N_{\delta o}$ are the number of radial points within the mechanical boundary layer at the inner and
outer boundary, respectively.
$\ell_{\rm max} = m_{\rm max}$ is the maximum degree and order of spherical harmonic expansion. 
Definitions of the Ekman number and modified Rayleigh number are given in (\ref{eq:NonDimNumbers}).
The amplitude of the heterogeneity in outer boundary heat flux is defined in (\ref{eq:qstardef}); $q_{\rm L}^\star = 0$ are homogeneous cases,
$Tq_{\rm L}^\star$ are cases with a pattern of heat flux derived from mantle tomography, $Hq_{\rm L}^\star$ are cases with
a hemispheric ($Y_1^1$) pattern.
$\Nus_{\rm T}$ is given by (\ref{eq:NuDef}). The Reynolds number is determined by (\ref{eq:PeReUDef})
and hence the kinetic energy integral (\ref{eq:KEdef}).
 $\Rey_{\rm pol}$ is found by retaining only the poloidal component of velocity (recall \ref{eq:VelTorPolDecomp})
 in the kinetic energy integral.
 $\Rey_{\rm zon}$ is found by retaining only the $m=0$ components from the spherical harmonic expansion of 
 the toroidal component of velocity in the kinetic energy integral. $P$ is the time average of the buoyancy production throughout
 the shell, $\epsilon_U$ is the time average of the viscous dissipation throughout the shell.

\begin{table}
  \begin{center}
\def~{\hphantom{0}}
 \begin{tabular}{cccccccccccc}
  Boundary  & $\widetilde{\Ray}$ & $\Nus_{\rm T}$ & $\Rey$ & $\Rey_{\rm pol}$ & $\Rey_{\rm zon}$  & $P$ & $\epsilon_{U}$ & $N_r$ & $\ell_{\rm max}$ & $N_{\delta i}$ & $N_{\delta o}$ \\[3pt]

 $q^\star_{\rm L}=0$ &         30 &       1.19 &       15.0 &        5.7 &        2.2 &  4.78021e+05 &  4.78022e+05 &         64 &         48 &         10 &         26 \\ 
 $q^\star_{\rm L}=0$ &         90 &       1.62 &       38.8 &       14.9 &       13.9 &  4.09815e+06 &  4.10101e+06 &         64 &         64 &         10 &         10 \\ 
 $q^\star_{\rm L}=0$ &        100 &       1.71 &       40.4 &       15.8 &       14.0 &  4.24413e+06 &  4.25138e+06 &         60 &         48 &          9 &         19 \\ 
 $q^\star_{\rm L}=0$ &        150 &       1.89 &       50.1 &       21.5 &       17.0 &  7.52380e+06 &  7.54178e+06 &         80 &         64 &         12 &         12 \\ 
 $q^\star_{\rm L}=0$ &        225 &       2.16 &       64.0 &       29.5 &       21.7 &  1.58309e+07 &  1.58508e+07 &         80 &         92 &         12 &         11 \\ 
 $q^\star_{\rm L}=0$ &        550 &       3.36 &       96.9 &       51.6 &       27.2 &  5.10257e+07 &  5.10242e+07 &         80 &         92 &         10 &         11 \\ 
 $q^\star_{\rm L}=0$ &        900 &       4.27 &      124.3 &       68.8 &       36.0 &  9.23021e+07 &  9.23000e+07 &         80 &         92 &          9 &         10 \\ 
 $q^\star_{\rm L}=0$ &       2000 &       5.92 &      193.4 &      107.1 &       70.2 &  2.24262e+08 &  2.24184e+08 &         96 &         96 &         12 &         13 \\ 
 $q^\star_{\rm L}=0$ &       4000 &       7.44 &      276.3 &      152.9 &      113.4 &  4.67205e+08 &  4.66904e+08 &         96 &         96 &         12 &         13 \\ 
 $q^\star_{\rm L}=0$ &       6000 &       8.35 &      332.8 &      189.0 &      129.4 &  7.08226e+08 &  7.08971e+08 &        128 &        128 &         15 &         18 \\ 
 $q^\star_{\rm L}=0$ &      13000 &      10.16 &      483.1 &      273.3 &      215.0 &  1.56626e+09 &  1.56477e+09 &        128 &        128 &         16 &         18 \\[3pt] 
 
 $Tq^\star_{\rm L}=2.3$        &         30 &       1.18 &       14.6 &        5.5 &        3.0 &  4.53139e+05 &  4.53161e+05 &         64 &         64 &         10 &         25 \\ 
 $Tq^\star_{\rm L}=2.3$        &         90 &       1.56 &       36.2 &       14.0 &       14.7 &  3.79416e+06 &  3.79750e+06 &         64 &         64 &         10 &         10 \\ 
 $Tq^\star_{\rm L}=2.3$        &        225 &       2.15 &       61.7 &       28.6 &       23.3 &  1.58478e+07 &  1.58566e+07 &         80 &         92 &         12 &         11 \\ 
$Tq^\star_{\rm L}=2.3$        &        550 &       3.38 &       93.3 &       50.7 &       29.3 &  5.11910e+07 &  5.12066e+07 &         80 &         92 &         10 &         10 \\ 
 $Tq^\star_{\rm L}=2.3$        &        900 &       4.33 &      125.3 &       68.8 &       33.6 &  9.28297e+07 &  9.28697e+07 &         80 &         92 &         10 &         10 \\ 
 $Tq^\star_{\rm L}=2.3$        &       2000 &       5.97 &      190.3 &      107.6 &       49.9 &  2.24701e+08 &  2.24704e+08 &         96 &         96 &         12 &         13 \\ 
$Tq^\star_{\rm L}=2.3$        &       4000 &       7.47 &      265.7 &      153.9 &       78.0 &  4.66833e+08 &  4.66660e+08 &         96 &         96 &         12 &         13 \\ 
 $Tq^\star_{\rm L}=2.3$        &       6000 &       8.41 &      322.3 &      186.7 &      107.3 &  7.11691e+08 &  7.11134e+08 &        128 &        128 &         15 &         17 \\ 
 $Tq^\star_{\rm L}=2.3$        &      13000 &      10.32 &      459.7 &      276.4 &      132.8 &  1.57402e+09 &  1.57212e+09 &        128 &        128 &         15 &         18 \\[3pt] 
 
 $Tq^\star_{\rm L}=5.0$        &         30 &       1.20 &       16.0 &        5.9 &        4.0 &  4.88902e+05 &  4.88902e+05 &         64 &         64 &         10 &         24 \\ 
 $Tq^\star_{\rm L}=5.0$        &         90 &       1.56 &       37.2 &       13.8 &       17.1 &  4.08174e+06 &  4.08853e+06 &         64 &         64 &         19 &         10 \\ 
 $Tq^\star_{\rm L}=5.0$        &        225 &       2.16 &       57.3 &       28.0 &       25.3 &  1.61914e+07 &  1.61994e+07 &         80 &         92 &         10 &         11 \\ 
$Tq^\star_{\rm L}=5.0$        &        550 &       3.57 &       93.8 &       50.8 &       33.4 &  5.40076e+07 &  5.40223e+07 &         80 &         92 &         10 &         10 \\ 
 $Tq^\star_{\rm L}=5.0$        &        900 &       4.59 &      121.9 &       68.3 &       38.7 &  9.67341e+07 &  9.67580e+07 &         80 &         92 &         10 &         10 \\ 
 $Tq^\star_{\rm L}=5.0$        &       2000 &       6.43 &      184.8 &      106.3 &       54.7 &  2.33029e+08 &  2.33003e+08 &         96 &         96 &         12 &         13 \\ 
 $Tq^\star_{\rm L}=5.0$        &       4000 &       8.21 &      261.0 &      153.7 &       67.4 &  4.83518e+08 &  4.83796e+08 &         96 &         96 &         12 &         13 \\ 
 $Tq^\star_{\rm L}=5.0$        &       6000 &       9.24 &      316.8 &      188.8 &       82.2 &  7.33091e+08 &  7.33801e+08 &        128 &        128 &         15 &         17 \\ 
$Tq^\star_{\rm L}=5.0$        &      13000 &      11.28 &      456.5 &      275.5 &      122.9 &  1.60985e+09 &  1.61080e+09 &        128 &        128 &         15 &         18 \\[3pt] 
 
 $Hq^\star_{\rm L}=2.3$        &         30 &       1.19 &       13.8 &        5.9 &        1.5 &  4.77130e+05 &  4.77107e+05 &         64 &         48 &         10 &         10 \\ 
 $Hq^\star_{\rm L}=2.3$        &         90 &       1.66 &       40.3 &       16.0 &       14.0 &  4.70090e+06 &  4.70703e+06 &         64 &         64 &         10 &         10 \\ 
 $Hq^\star_{\rm L}=2.3$        &        225 &       2.17 &       62.8 &       29.5 &       22.9 &  1.61102e+07 &  1.61326e+07 &         80 &         92 &         12 &         11 \\ 
 $Hq^\star_{\rm L}=2.3$        &        550 &       3.47 &      103.0 &       52.6 &       32.7 &  5.24670e+07 &  5.25358e+07 &         80 &         92 &         10 &         11 \\ 
 $Hq^\star_{\rm L}=2.3$        &        900 &       4.41 &      134.5 &       70.5 &       42.7 &  9.39105e+07 &  9.39404e+07 &         80 &         92 &         10 &         11 \\ 
 $Hq^\star_{\rm L}=2.3$        &       2000 &       6.20 &      204.9 &      110.1 &       66.0 &  2.27285e+08 &  2.27356e+08 &         96 &         96 &         12 &         13 \\ 
 $Hq^\star_{\rm L}=2.3$        &       4000 &       7.78 &      287.5 &      157.9 &       98.2 &  4.69832e+08 &  4.70235e+08 &         96 &         96 &         12 &         13 \\ 
 $Hq^\star_{\rm L}=2.3$        &       6000 &       8.76 &      348.5 &      194.7 &      118.2 &  7.15230e+08 &  7.15039e+08 &        128 &        128 &         16 &         18 \\ 
 $Hq^\star_{\rm L}=2.3$        &      13000 &      10.76 &      502.3 &      283.2 &      169.8 &  1.58059e+09 &  1.57940e+09 &        128 &        128 &         16 &         18 \\[3pt] 
 
 $Hq^\star_{\rm L}=5.0$        &         30 &       1.42 &       22.5 &        8.8 &        5.7 &  1.30797e+06 &  1.30797e+06 &         64 &         48 &         10 &         17 \\ 
 $Hq^\star_{\rm L}=5.0$        &         90 &       1.73 &       42.7 &       17.1 &       16.6 &  5.64898e+06 &  5.66254e+06 &         64 &         64 &         10 &         10 \\ 
 $Hq^\star_{\rm L}=5.0$        &        225 &       2.33 &       67.3 &       31.5 &       26.0 &  1.86103e+07 &  1.86548e+07 &         80 &         92 &         12 &         11 \\ 
 $Hq^\star_{\rm L}=5.0$        &        550 &       3.78 &      106.7 &       55.2 &       37.0 &  5.75056e+07 &  5.75666e+07 &         80 &         92 &         10 &         11 \\ 
 $Hq^\star_{\rm L}=5.0$        &        900 &       5.00 &      141.3 &       74.0 &       46.0 &  1.02083e+08 &  1.02217e+08 &         80 &         92 &         10 &         11 \\ 
 $Hq^\star_{\rm L}=5.0$        &       2000 &       7.10 &      211.8 &      114.0 &       67.6 &  2.41234e+08 &  2.41658e+08 &         96 &         96 &         12 &         13 \\ 
 $Hq^\star_{\rm L}=5.0$        &       4000 &       8.92 &      296.3 &      162.3 &      102.9 &  4.94373e+08 &  4.94661e+08 &        128 &        128 &         16 &         19 \\ 
 $Hq^\star_{\rm L}=5.0$        &       6000 &      10.07 &      359.5 &      197.8 &      130.4 &  7.48688e+08 &  7.49483e+08 &        128 &        128 &         16 &         19 \\ 
 $Hq^\star_{\rm L}=5.0$        &      13000 &      12.25 &      509.5 &      286.1 &      187.4 &  1.64536e+09 &  1.64421e+09 &        128 &        128 &         16 &         20 \\ 
\end{tabular}

  \caption{Summary of all runs for $\Ek = 10^{-4}$.}
  \label{tab:fullresultsE-4}
  \end{center}
\end{table}

\begin{table}
  \begin{center}
\def~{\hphantom{0}}
  \begin{tabular}{cccccccccccc}
Boundary  & $\widetilde{\Ray}$ & $\Nus_{\rm T}$ & $\Rey$ & $\Rey_{\rm pol}$ & $\Rey_{\rm zon}$  & $P$ & $\epsilon_{U}$ & $N_r$ & $\ell_{\rm max}$ & $N_{\delta i}$ & $N_{\delta o}$ \\[3pt]
  
    $q^\star_{\rm L}=0$ &         30 &       1.03 &        7.1 &        3.4 &        0.7 &  4.30198e+05 &  4.30198e+05 &         80 &         64 &         28 &          5 \\ 
   $q^\star_{\rm L}=0$ &         90 &       1.25 &       38.5 &       16.5 &       14.5 &  1.34837e+07 &  1.34839e+07 &         90 &         80 &         22 &          7 \\ 
   $q^\star_{\rm L}=0$ &        150 &       1.43 &       65.7 &       27.0 &       21.6 &  3.98206e+07 &  3.98214e+07 &         90 &         80 &         22 &          8 \\ 
   $q^\star_{\rm L}=0$ &        225 &       1.58 &       86.2 &       35.3 &       34.1 &  8.12987e+07 &  8.13030e+07 &         90 &        128 &          9 &          7 \\ 
   $q^\star_{\rm L}=0$ &        550 &       2.29 &      150.2 &       71.8 &       51.1 &  3.74397e+08 &  3.74562e+08 &         90 &        128 &          7 &          8 \\ 
   $q^\star_{\rm L}=0$ &        900 &       3.13 &      199.0 &      101.2 &       62.9 &  7.58361e+08 &  7.58773e+08 &         90 &        128 &          7 &          8 \\ 
   $q^\star_{\rm L}=0$ &       1200 &       3.75 &      234.2 &      121.7 &       69.3 &  1.09936e+09 &  1.09929e+09 &         90 &        128 &          7 &          8 \\ 
   $q^\star_{\rm L}=0$ &       2000 &       5.14 &      316.1 &      167.4 &       95.0 &  2.06032e+09 &  2.06007e+09 &        128 &        144 &          9 &         10 \\ 
   $q^\star_{\rm L}=0$ &       2500 &       5.88 &      361.2 &      190.4 &      121.1 &  2.67403e+09 &  2.67356e+09 &        128 &        144 &          9 &         10 \\ 
  $q^\star_{\rm L}=0$ &       4000 &       7.67 &      472.3 &      249.8 &      170.4 &  4.52798e+09 &  4.52712e+09 &        192 &        192 &         14 &         14 \\ 
   $q^\star_{\rm L}=0$ &       6000 &       9.47 &      598.3 &      310.3 &      259.7 &  7.02322e+09 &  7.02222e+09 &        192 &        192 &         14 &         14 \\ 
   $q^\star_{\rm L}=0$ &      13000 &      13.29 &      927.8 &      456.6 &      512.4 &  1.58024e+10 &  1.57922e+10 &        256 &        256 &         16 &         14 \\[3pt] 
 
 $Tq^\star_{\rm L}=2.3$        &         30 &       1.02 &        6.5 &        2.8 &        2.3 &  3.64894e+05 &  3.64895e+05 &        128 &         96 &         28 &          7 \\ 
 $Tq^\star_{\rm L}=2.3$        &         90 &       1.24 &       38.3 &       15.7 &       15.1 &  1.26866e+07 &  1.26866e+07 &        128 &         96 &         33 &         10 \\ 
$Tq^\star_{\rm L}=2.3$        &        225 &       1.60 &       89.4 &       36.3 &       34.5 &  8.45382e+07 &  8.45432e+07 &        128 &         96 &         33 &         10 \\ 
 $Tq^\star_{\rm L}=2.3$        &        550 &       2.30 &      146.4 &       70.6 &       52.2 &  3.75773e+08 &  3.75913e+08 &         90 &        128 &          7 &          8 \\ 
 $Tq^\star_{\rm L}=2.3$        &        900 &       3.14 &      195.0 &       99.4 &       62.7 &  7.62351e+08 &  7.62548e+08 &         90 &        128 &          7 &          8 \\ 
 $Tq^\star_{\rm L}=2.3$        &       1200 &       3.77 &      229.3 &      120.1 &       69.7 &  1.10618e+09 &  1.10655e+09 &         90 &        128 &          7 &          8 \\ 
 $Tq^\star_{\rm L}=2.3$        &       2000 &       5.17 &      310.2 &      164.3 &       94.4 &  2.06876e+09 &  2.06925e+09 &        128 &        144 &          9 &         10 \\ 
 $Tq^\star_{\rm L}=2.3$        &       2500 &       5.89 &      351.1 &      187.0 &      110.2 &  2.68093e+09 &  2.68059e+09 &        128 &        144 &          9 &         10 \\ 
$Tq^\star_{\rm L}=2.3$        &       4000 &       7.65 &      456.4 &      244.0 &      161.8 &  4.52501e+09 &  4.52599e+09 &        192 &        192 &         14 &         14 \\ 
$Tq^\star_{\rm L}=2.3$        &       6000 &       9.37 &      564.1 &      305.0 &      201.2 &  7.01606e+09 &  7.01647e+09 &        192 &        192 &         14 &         14 \\ 
 $Tq^\star_{\rm L}=2.3$        &      13000 &      13.13 &      861.3 &      452.2 &      402.3 &  1.57709e+10 &  1.57698e+10 &        192 &        192 &         14 &         14 \\[3pt] 
 
 $Tq^\star_{\rm L}=5.0$        &         30 &       1.01 &        7.9 &        2.6 &        4.3 &  4.87080e+05 &  4.87079e+05 &        128 &         96 &         52 &         12 \\ 
 $Tq^\star_{\rm L}=5.0$        &         90 &       1.27 &       41.6 &       16.9 &       13.5 &  1.48061e+07 &  1.48063e+07 &         90 &         96 &         23 &         16 \\ 
 $Tq^\star_{\rm L}=5.0$        &        225 &       1.59 &       90.1 &       35.4 &       38.6 &  8.71763e+07 &  8.72294e+07 &        128 &         96 &         31 &         10 \\ 
 $Tq^\star_{\rm L}=5.0$        &        550 &       2.38 &      145.6 &       69.9 &       59.7 &  3.95231e+08 &  3.95306e+08 &         90 &        128 &          7 &          8 \\ 
 $Tq^\star_{\rm L}=5.0$        &        900 &       3.27 &      190.3 &       98.4 &       66.7 &  7.96897e+08 &  7.97092e+08 &         90 &        128 &          7 &          8 \\ 
 $Tq^\star_{\rm L}=5.0$        &       1200 &       3.92 &      223.2 &      118.0 &       73.5 &  1.15361e+09 &  1.15369e+09 &         90 &        128 &          7 &          8 \\ 
 $Tq^\star_{\rm L}=5.0$        &       2000 &       5.39 &      296.0 &      160.7 &       91.5 &  2.13375e+09 &  2.13370e+09 &        128 &        144 &          9 &         10 \\ 
$Tq^\star_{\rm L}=5.0$        &       2500 &       6.16 &      336.6 &      183.8 &      101.5 &  2.76352e+09 &  2.76337e+09 &        128 &        144 &          9 &         10 \\ 
$Tq^\star_{\rm L}=5.0$        &       4000 &       8.08 &      433.5 &      240.8 &      129.5 &  4.65028e+09 &  4.65196e+09 &        192 &        192 &         14 &         14 \\ 
 $Tq^\star_{\rm L}=5.0$        &       6000 &      10.00 &      540.7 &      300.1 &      179.4 &  7.19371e+09 &  7.19458e+09 &        192 &        192 &         14 &         14 \\ 
 $Tq^\star_{\rm L}=5.0$        &      13000 &      14.22 &      812.1 &      446.4 &      327.4 &  1.61015e+10 &  1.61076e+10 &        192 &        192 &         14 &         14 \\[3pt] 
 
 $Hq^\star_{\rm L}=2.3$        &         30 &       1.04 &       10.7 &        4.7 &        1.1 &  1.04198e+06 &  1.04198e+06 &         80 &         64 &         34 &         12 \\ 
 $Hq^\star_{\rm L}=2.3$        &         90 &       1.31 &       44.6 &       19.4 &       12.5 &  1.90383e+07 &  1.90388e+07 &         90 &         80 &         34 &          8 \\ 
 $Hq^\star_{\rm L}=2.3$        &        225 &       1.62 &       91.8 &       38.0 &       33.3 &  9.29168e+07 &  9.30683e+07 &         90 &        128 &         23 &          8 \\ 
 $Hq^\star_{\rm L}=2.3$        &        550 &       2.33 &      154.5 &       73.7 &       53.5 &  3.90540e+08 &  3.90853e+08 &         90 &        128 &          7 &          8 \\ 
 $Hq^\star_{\rm L}=2.3$        &        900 &       3.19 &      205.7 &      103.3 &       66.7 &  7.80866e+08 &  7.81356e+08 &         90 &        128 &          7 &          8 \\ 
 $Hq^\star_{\rm L}=2.3$        &       2000 &       5.33 &      332.3 &      170.9 &      103.1 &  2.11004e+09 &  2.11030e+09 &        128 &        144 &         10 &         10 \\ 
 $Hq^\star_{\rm L}=2.3$        &       4000 &       8.00 &      493.1 &      253.9 &      191.6 &  4.59532e+09 &  4.59673e+09 &        192 &        192 &         14 &         14 \\ 
 $Hq^\star_{\rm L}=2.3$        &       6000 &       9.81 &      616.3 &      315.6 &      269.7 &  7.10168e+09 &  7.10107e+09 &        192 &        192 &         14 &         14 \\ 
 $Hq^\star_{\rm L}=2.3$        &      13000 &      13.74 &      938.2 &      466.5 &      492.8 &  1.59043e+10 &  1.59169e+10 &        256 &        256 &         20 &         20 \\[3pt] 
 
 $Hq^\star_{\rm L}=5.0$        &         30 &       1.08 &       17.1 &        7.2 &        2.0 &  2.86108e+06 &  2.86107e+06 &         80 &         64 &         36 &          7 \\ 
 $Hq^\star_{\rm L}=5.0$        &         90 &       1.48 &       62.8 &       25.7 &       22.8 &  4.47602e+07 &  4.47520e+07 &         96 &         96 &         25 &         10 \\ 
  $Hq^\star_{\rm L}=5.0$        &        225 &       1.68 &       97.8 &       40.8 &       36.9 &  1.12445e+08 &  1.12655e+08 &         90 &        128 &         24 &          8 \\ 
 $Hq^\star_{\rm L}=5.0$        &        550 &       2.42 &      160.5 &       77.1 &       59.5 &  4.29394e+08 &  4.29842e+08 &         90 &        128 &          7 &          8 \\ 
 $Hq^\star_{\rm L}=5.0$        &        900 &       3.37 &      213.5 &      107.3 &       72.5 &  8.41615e+08 &  8.42580e+08 &         90 &        128 &          7 &          8 \\ 
 $Hq^\star_{\rm L}=5.0$        &       2000 &       5.77 &      339.3 &      175.2 &      101.6 &  2.22759e+09 &  2.23006e+09 &        128 &        144 &         10 &         10 \\ 
 $Hq^\star_{\rm L}=5.0$        &       4000 &       9.01 &      498.5 &      259.7 &      173.8 &  4.84436e+09 &  4.85116e+09 &        192 &        192 &         14 &         15 \\ 
 $Hq^\star_{\rm L}=5.0$        &       6000 &      11.39 &      622.6 &      321.5 &      246.0 &  7.45281e+09 &  7.46115e+09 &        192 &        192 &         14 &         15 \\ 
 $Hq^\star_{\rm L}=5.0$        &      13000 &      16.49 &      958.4 &      488.9 &      437.4 &  1.69246e+10 &  1.70334e+10 &        256 &        256 &         21 &         20 \\ 

\end{tabular}

  \caption{Summary of all runs with $\Ek = 10^{-5}$.}
  \label{tab:fullresultsE-5}
  \end{center}
\end{table}

\begin{table}
  \begin{center}
\def~{\hphantom{0}}
  \begin{tabular}{cccccccccccc}
 Boundary  & $\widetilde{\Ray}$ & $\Nus_{\rm T}$ & $\Rey$ & $\Rey_{\rm pol}$ & $\Rey_{\rm zon}$  & $P$ & $\epsilon_{U}$ & $N_r$ & $\ell_{\rm max}$ & $N_{\delta i}$ & $N_{\delta o}$ \\[3pt]
  
        $q^\star_{\rm L}=0$ &       2000 &       3.19 &      504.1 &      251.6 &      139.0 &  1.61586e+10 &  1.61648e+10 &        192 &        192 &         11 &         10 \\ 
        $q^\star_{\rm L}=0$ &       6000 &       6.51 &      951.9 &      488.2 &      308.6 &  6.00882e+10 &  6.02028e+10 &        224 &        224 &         12 &         12 \\ 
        $q^\star_{\rm L}=0$ &      18000 &      12.92 &     1670.0 &      849.7 &      689.4 &  1.70352e+11 &  1.71287e+11 &        320 &        320 &         16 &         15 \\[3pt] 
 
 $Tq^\star_{\rm L}=2.3$        &       2000 &       3.19 &      489.6 &      245.7 &      140.2 &  1.62591e+10 &  1.62616e+10 &        192 &        192 &         11 &         10 \\ 
 $Tq^\star_{\rm L}=2.3$        &       6000 &       6.50 &      926.9 &      482.1 &      282.8 &    6.28045e+10 &  6.28064e+10 &        224 &        224 &         12 &         12 \\ 
 $Tq^\star_{\rm L}=2.3$        &      18000 &      12.87 &     1768.5 &      893.8 &      790.8 &    2.10229e+11 &  2.10222e+11 &        320 &        320 &         16 &         15 \\[3pt] 
 
  $Tq^\star_{\rm L}=5.0$        &       2000 &       3.26 &      478.2 &      241.7 &      146.7 &  1.68536e+10 &  1.68587e+10 &        192 &        192 &         11 &         10 \\ 
  $Tq^\star_{\rm L}=5.0$        &       6000 &       6.75 &      881.6 &      460.8 &      274.3 &  6.07211e+10 &  6.08708e+10 &        224 &        224 &         11 &         12 \\ 
  $Tq^\star_{\rm L}=5.0$        &      18000 &      13.55 &     1698.9 &      873.6 &      693.6 &  2.08698e+11 &  2.07256e+11 &        320 &        320 &         16 &         15 \\

  \end{tabular}

  \caption{Summary of all runs with $\Ek = 10^{-6}$.}
  \label{tab:fullresultsE-6}
  \end{center}
\end{table}

\newpage

\bibliographystyle{jfm}
\bibliography{MD17}

\end{document}